\documentclass[%
 reprint,
showpacs,preprintnumbers,
 amsmath,amssymb,
 aps,
pra,
]{revtex4-1}

\usepackage{graphicx}
\usepackage{dcolumn}
\usepackage {subfigure}
\input {psfig.sty}
\usepackage{bm}

\usepackage{amsmath}
\usepackage{amsthm}

\begin{document}

\preprint{APS/123-QED}

\title{Post-processing of the oblivious key in quantum private queries}

\author{Fei Gao}
 \altaffiliation[Also at ]{State Key Laboratory of Integrated Service Networks, Xidian University, Xi'an 710071, China}
 \email{gaofei\_bupt@hotmail.com}
\author{Bin Liu}
 \email{lbhitmanbl@gmail.com}
\author{Wei Huang}
\author{Qiao-Yan Wen}
\affiliation{
 State Key Laboratory of Networking and Switching Technology, Beijing University of Posts and Telecommunications, Beijing, 100876, China
}

\date{\today}

\begin{abstract}
Quantum private query (QPQ) is a kind of quantum protocols to protect both users' privacy in their communication. There is an interesting example, that is, Alice wants to buy one item from Bob's database, which is composed of a quantity of valuable messages. QPQ protocol is the communication procedure ensuring that Alice can get only one item from Bob, and at the same time, Bob cannot know which one was taken by Alice. Owing to its practicability, quantum-key-distribution-based QPQ has draw much attention in recent years. However, the post-processing of the key in such protocols, called oblivious key, remains far from being satisfactorily known. Especially, the error correction method for such special key is still missing. Here we focus on the post-processing of the oblivious key, including both dilution and error correction. On the one hand, we demonstrate that the previous dilution method, which greatly reduces the communication complexity, will bring Alice the chance to illegally obtain much additional information about Bob's database. Simulations show that by very limited queries Alice can obtain the whole database. On the other hand, we present an effective error-correction method for the oblivious key, which completes its post-processing and makes such QPQ more practical.
\end{abstract}

\maketitle


\section{\label{sec:1}Introduction}

In some cryptographic communications, we need not only protect the security of the transmitted message against eavesdropping from an outside adversary, but also the communicators' individual privacy against each other. An interesting example is the problem of private user queries to a database, where Alice bought an item in Bob's database, which is composed of a quantity of valuable messages, and wants to get it from Bob (given that Alice knows the address of this item in the database). Here Bob will worry about that Alice can obtain more items from his database when she gets the bought one, and Alice might not want Bob know which item she is querying. Symmetrically private information retrieval (SPIR) \cite{SPIR} protocols are designed for such circumstance. It ensures the privacies of both Alice and Bob. That is, after the communication, Alice can correctly obtain at most one item (i.e. the one who bought) from Bob's database and simultaneously, Bob does not know the address Alice has retrieved.

As we know, the security of most classical cryptosystems is based on the assumptions of computational complexity, and they might be broken by the strong ability of some advanced algorithms like quantum computation \cite{Shor,Grover}. Fortunately, this difficulty can be overcome by quantum cryptography \cite{BB84,GRTZ02}, where the security is assured by physical principles. Owing to its higher security, quantum cryptography has attracted a great deal of attention now.

Quantum Private Queries (QPQ) is the quantum scheme for SPIR problem.  In fact, as a kind of protocols for two party quantum secure computations, the task of SPIR cannot be achieved ideally even in its quantum version \cite{Lo1997}. More practically, the requirements are generally relaxed into that Alice can elicit sufficiently little content of the database, and Bob's attack will be discovered, with a certain probability, by Alice if he tries to obtain the address Alice is retrieving.

In 2008 V. Giovannetti et al proposed the first QPQ protocol \cite{QPQ08,QPQproof,QPQexp}, where the database is represented by a unitary operation (i.e. oracle operation) and it is performed on the two coming query states. In 2011 L. Olejnik presented an improved protocol where only one query state is necessary \cite{QPIR11}. Compared with the previous (classical) SPIR schemes, the above two protocols display an exponential reduction in both communication complexity and running-time computational complexity.

Though the above two protocols exhibit significant advantages in theory, they are difficult to implement because when large database is concerned the dimension of the oracle operation will be very high. To solve this problem, M. Jakobi et al gave a new QPQ protocol (J-protocol) based on quantum key distribution (QKD) \cite{QPQ11}. In this protocol SARG04 QKD scheme \cite{SARG04} is utilized to distribute an oblivious key between Alice and Bob, which satisfies the following three requirements.
\begin{itemize}
\item[R1.] Bob knows every bit in the key.\vspace{-2.5mm}
\item[R2.] Alice knows every key bit with a certain probability (e.g. 0.25 for an honest Alice in J-protocol).\vspace{-2.5mm}
\item[R3.] Bob does not know which key bits are known by Alice.
\end{itemize}
Then the post-processing of the key is needed, which changes the oblivious key so that it satisfies the above requirements (R1), (R3), and a modified (R2), that is,
\begin{itemize}
\item[R2'.] Alice knows only several bits in the key (note that the ideal case is Alice knows just one bit, but to increase the success probability of the communication the bits obtained by Alice are generally little more than one, e.g. 2-7 on average).
\end{itemize}
Here we call the key before and after post-processing the Raw Oblivious Key (ROK) and the Final Oblivious key (FOK), respectively. Obviously it is a ``dilution process'' from ROK to FOK. Afterwards, if Alice knows the $j$-th bit in the FOK and wants to retrieve the $i$-th item in Bob's database, she declares a shift value $s=j-i$ so that Bob can shift his FOK by $s$. At last Bob encrypts his database by the shifted FOK (for simplicity, it is generally assumed that the content of each item in the database is just a bit, and one key bit can encrypt one item by ont-time pad), and sends the whole encrypted database to Alice. Thus Alice can correctly decrypt the item she wanted by her known key bit. More importantly, both users' privacy are successfully protected.

Compared with previous QPQ protocols, J-protocol is based on QKD, the most practical application of quantum information technology, and consequently it is easy to be realized. More concretely, it can be easily generalized to large database, and it is loss tolerant. Therefore, as a practical model of QPQ, QKD-based QPQ is very attractive and has become a studying point. Different manners to distribute the ROK were given by some of us in Refs. \cite{OE12,PRA13}, and different methods for post-processing were presented in Refs. \cite{PJ13,SZM12}.

Here we focus on the post-processing of the oblivious key in QPQ. In Ref. \cite{QPQ11} M. Jakobi et al gave a $kN-N$ method, that is, it transforms a ROK with length $kN$ into a $N$-bit FOK (here $N$ is the total number of items in the database and $k$ is a parameter which is an integer greater than 1). Afterwards M. V. Panduranga Rao et al proposed two improved methods in Ref. \cite{PJ13}. One is $N-N$, and the other is $rM-N$, where $r,M$ are integers satisfying $rM\ll N$. Obviously these two methods would greatly reduce the communication complexity.

However, in our opinion, two important problems about the post-processing of the oblivious key need be further studied.

(1) As pointed by the authors of Ref. \cite{PJ13}, the reduction in communication complexity in their methods comes at the cost that parity information about some key bits in FOK is easier to obtain for Alice. So, the question is, what kind of influence this fact (that is, the parity information were illegally leaked to Alice) brings on the database security?

(2) In a practical realization, noise exists in the channel and there are always errors in the shared key between Alice and Bob. An error bit in Alice's FOK implies that Alice would pay her money and get back a wrong content from Bob, which is obviously unfair for Alice. Moreover, when she finds the bought content is error, Alice believes Bob is cheating, and consequently Bob's reputation will become very bad. Therefore, just like in QKD, error correction is necessary in the post-processing. In fact, dilution and error correction compose the whole post-processing of oblivious key in QPQ (note that privacy amplification, another important part of post-processing in QKD, is unnecessary here because its function will be achieved by the dilution procedure). But performing error correction on oblivious key is difficult. This is because, to correct errors, Bob has to declare additional information about the key to Alice, which would bring the chance for Alice to illegally get much more key bits than expected. M. Jakobi et al listed error correction as an open question in QPQ \cite{QPQ11}, and till now all previous post-processing methods just pay attention to ``dilution'', while error correction is still awaited.

In this paper we demonstrate that M. V. Panduranga Rao et al's dilution methods would result in insecurity for Bob's database. Simulations show that Alice can obtain much more items (even the whole database) than expected if she execute multiple but limited (averagely 53.4 for $N=10^4$) queries. In other words, buying about 53 items means obtaining the whole database with $10^4$ items in total! Furthermore, we propose an effective error-correction method for the oblivious key, which completes its post-processing and makes such QPQ protocol more practical for a real noisy channel.

The rest of this paper is organized as follows. In Sec. II we give a brief review of previous three dilution methods, and analysis the insecurity brought by the $N-N$ method and the $rM-N$ one in Sec. III. The scheme of post-processing with error-correction is presented In Sec. IV and Sec.V is our conclusion.

\section{Previous dilution methods}

In this section we will give brief sketches of previous dilution methods, including the $kN-N$ one given by M. Jakobi et al \cite{QPQ11}, and the improved ones, i.e. $N-N$ and $rM-N$ ones, given by M. V. Panduranga Rao et al \cite{PJ13}. In all these methods it is assumed that Alice and Bob have shared a ROK $O^R$ satisfying the above requirements R1-R3. The aim is to dilute $O^R$ into the FOK $O^F$ satisfying R1,R2', and R3.

\subsection{The $kN-N$ method}
For simplicity, the $kN$-bit ROK $O^R$ can be denoted as ${O^R_1O^R_2...O^R_{kN}}$, and the $N$-bit FOK $O^F$ as ${O^F_1O^F_2...O^F_N}$. Here every $O^R_i (1\leq i\leq kN)$ or $O^F_i (1\leq i\leq N)$ represents a key bit. In this method the relation between $O^R$ and $O^F$ is
\begin{eqnarray}
O^F_i=\bigoplus_{j=0}^{k-1} O^R_{i+jN}, 1\leq i\leq N,
\label{eq:one}
\end{eqnarray}
where $\bigoplus$ denotes the addition modulo 2.

For example, $N=12$, $k=2$, and the ROK is
\begin{eqnarray}
0~1~1~0,~0~1~0~0,~0~1~1~1\nonumber\\
0~0~1~1,~0~1~0~1,~1~0~0~1
\end{eqnarray}
for Bob, while
\begin{eqnarray}
?~1~?~?,~0~?~?~?,~?~1~?~?\nonumber\\
0~?~?~?,~?~1~?~?,~?~0~?~?
\end{eqnarray}
for Alice (that is, Alice only knows the 2nd, 5th, 10th, 13rd, 18th, 22nd bits in this ROK).
Then after the dilution, the FOK is
\begin{eqnarray}
0~1~0~1,~0~0~0~1,~1~1~1~0
\end{eqnarray}
for Bob, while
\begin{eqnarray}
?~?~?~?,~?~?~?~?,~?~1~?~?
\end{eqnarray}
for Alice. It's easy to see that the quantity of the known bits for Alice is reduced from 6 to 1.

\subsection{The $N-N$ method}
Similarly, the $N$-bit ROK $O^R$ can be denoted as ${O^R_1O^R_2...O^R_{N}}$, and the $N$-bit FOK $O^F$ as ${O^F_1O^F_2...O^F_N}$. In this $N-N$ method the relation between $O^R$ and $O^F$ is
\begin{eqnarray}
O^F_i=\bigoplus_{j=i}^{i+k-1\mod N} O^R_{j}, 1\leq i\leq N.
\end{eqnarray}

For example, $N=12$, $k=2$, and the ROK is
\begin{eqnarray}
0~1~1~0,~0~1~0~0,~0~1~1~1
\end{eqnarray}
for Bob, while
\begin{eqnarray}
?~?~?~0,~0~?~0~?,~?~?~?~?
\end{eqnarray}
for Alice.
Then after the dilution, the FOK is
\begin{eqnarray}
1~0~1~0,~1~1~0~0,~1~0~0~1
\end{eqnarray}
for Bob, while
\begin{eqnarray}
?~?~?~0,~?~?~?~?,~?~?~?~?
\end{eqnarray}
for Alice. It's easy to see that the quantity of the known bits for Alice is reduced from 3 to 1.

\subsection{The $rM-N$ method}\label{rm-n}
In this method, the $rM$-bit ROK $O^R$ is divided into $r$ sub-keys with the same length $M$, i.e. $O^{R_1},O^{R_2},...,O^{R_r}$. Then the following two steps are executed.

(1) \textbf{Sub-key extension}. For every sub-key $O^{R_i} (1\leq i\leq r)$, the parities (i.e. the sum modulo 2) of all possible combinations of $k$ out of these $M$ bits, listed by a certain order, compose a new key $\widetilde{O}^{R_i}$. In fact, there are $(^M_k)$ combinations of $k$ out of $M$, so the length of each new key $\widetilde{O}^{R_i}$ is $(^M_k)$, which is generally supposed to be equal to the database's capacity $N$.

(2) \textbf{Shift-addition}. To obtain the FOK $O^F$, the above $r$ keys $\widetilde{O}^{R_i}$ are combined bitwise with relative shifts $s_i$ Alice can freely choose, that is
\begin{eqnarray}
O^F_j=\bigoplus_{i=1}^r \widetilde{O}^{R_i}_{j+s_i}, 1\leq j\leq N,
\end{eqnarray}
where $O^F_j$ represents the $j$-th bit in $O^F$ and $\widetilde{O}^{R_i}_{j+s_i}$ is the $(j+s_i)$-th bit in $\widetilde{O}^{R_i}$.

The first step tries to reuse every bit in ROK to the most degree so that $M$, i.e. the communication complexity for each sub-key $O^{R_i}$, reaches the lowest value. But obviously it will result in the amount of the bits Alice knows in $\widetilde{O}^{R_i}$ is more than expected. Therefore, the second step is utilized to reduce Alice's knowledge in the final key. Because Alice can choose a shift for each $\widetilde{O}^{R_i}$, she will know at least 1 bit in the final key $O^F$ given that she knows at least 1 bit in each $\widetilde{O}^{R_i}$. The feature of shift-addition is that it can reduce Alice's known bits in the final key, while, at the same time, dose not increase the failure probability (that is, Alice knows no bit in the final key). Please sea Ref. \cite{PJ13} for details (the technique of shift-addition was also discussed in Ref. \cite{QPQ11}) .

\section{Security analysis on the improved dilution methods}

In all the three dilution methods the parity of $k$ ROK bits means a FOK bit. It is not difficult to see that in the improved dilution methods, i.e. $N-N$ and $rM-N$ ones, the bits in ROK are reused frequently so that the communication complexity is greatly reduced. More specifically, ROK with length $N$ or even $rM$, instead of the original $kN$, is enough for $N$-bit FOK. However, as pointed by the authors of Ref. \cite{PJ13}, the reduction in communication complexity in their methods comes at the cost that parity information about some key bits in FOK is easier to obtain for Alice. Taking the $N-N$ mothod as our example, though Alice does not know their particular values, Alice does know the parity of two adjacent FOK bits $O^F_i$ and $O^F_{i+1}$ if Alice knows two ROK bits $O^R_i$ and $O^R_{i+k}$ (obviously $O^F_i\oplus O^F_{i+1}=O^R_i\oplus O^R_{i+k}$). Similarly, if Alice also knows $O^R_{i+1}$ and $O^R_{i+k+1}$ she knows the parity of $O^F_{i+1}$ and $O^F_{i+2}$, and consequently knows all parities of any two bits in the adjacent-bit set $\{O^F_i,O^F_{i+1},O^F_{i+2}\}$. Here we say such sets, like the above $\{O^F_i,O^F_{i+1}\}$ and $\{O^F_i,O^F_{i+1},O^F_{i+2}\}$, are \textit{almost known} (for Alice) because in each of them Alice would know all of the bits if she gets any one in it.

Now, the question is, what kind of influence this fact (that is, the parity information were illegally leaked to Alice) brings on the database's security? Remember that, in the database, every item, i.e. one-bit secret message $m_i$, will be encrypted through $O^F_i$, getting the ciphertext $c_i=m_i\oplus O^F_i$, and then all the ciphertext bits will be transmitted to Alice. Therefore, the message sets such as $\{m_i,m_{i+1}\}$ and $\{m_i,m_{i+1},m_{i+2}\}$, which correspond to the above almost-known-final-key sets, are also almost known for Alice. It implies Alice can illegally obtain much more information about the secret messages in database than expected. If only one query is considered such information leakage seems trivial because Alice does not know more explicit messages than expected. However, if Alice (or equivalent, Alice and some other users who collude with her) executes multiple queries to Bob's database by buying multiple messages from him, the influence of this kind of information leakage will become serious.

In fact, in each query, apart from obtaining 1 or little more bits in FOK (equivalently, 1 or little more items in the database), Alice would also get additional information, that is, identifying some Almost Known Sets (AKS) in the database. With the increasing amount of the queries, such AKSs will become more and more, and then some of them might be combined together generating a larger set (for example, a previous AKS $\{m_i,m_{i+1}\}$ and a new AKS $\{m_{i+1},m_{i+2}\}$ will be combined together into a larger one $\{m_i,m_{i+1},m_{i+2}\}$, and previous two AKSs $\{m_i,m_{i+1}\}$ and $\{m_{i+2},m_{i+3},m_{i+4}\}$ will be linked together by a new AKS $\{m_{i+1},m_{i+2}\}$, becoming a larger one $\{m_i,m_{i+1},m_{i+2},m_{i+3},m_{i+4}\}$). If Alice legally obtains a bit in an AKS in one of the queries, all the items in this set will be \textit{lighted}, that is, all of them are explicitly known for Alice. It is not difficult to imagine that after a certain number of queries Alice will obtain all the database completely. That is to say, Alice might steal the whole database, which is full of valuable secret messages, by just buying limited items in it. Furthermore, Alice can choose an optimal shift, i.e. the shift on the FOK before encrypting the database, in each query so that she can obtain the whole database with less queries.

Indeed, Alice can obtain the whole database by enough queries (at most $N$) even though she does not utilize the information brought by those AKS. The key point is, would Alice, with the help of the illegal information, achieve her goal greatly more quickly than expected? The answer is yes. It means the leaked parity information about some key bits in FOK results in serious insecurity for Bob's database.

\subsection{Analysis on the $N-N$ method}
As described in Sec. \ref{sec:1}, Alice will know every ROK bit with probability $p=1/4$ if she is honest in the J-protocol. While if she prepares a quantum memory and executes individual unambiguous state discrimination (USD) measurement to attack, Alice can increase this probability to $p=1-\frac{1}{\sqrt{2}}\approx0.29$ \cite{QPQ11}. Then the probability of identifying a two-bit AKS such as $\{m_i,m_{i+1}\}$ is 1/16 and $3/2-\sqrt{2}$ in the above two conditions respectively. Now we do simulations for different parameters to see how much information Alice will illegally obtain from the leaked parity information, especially how many queries are needed on average for Alice to obtain the whole database.

In our simulation, the approach Alice chooses to attack by multiple queries is as follows.

(S1) Define the parameters $N,k$, and $p$. Here $N$ is the total number of items in Bob's database, the integer $k$ is a security parameter which is chosen so that the quantity of Alice's known bits in FOK $c=\frac{N}{4^k}$ is little more than 1 and the failure (that is, no bit is obtained by Alice) probability is small enough \cite{PJ13}, and $p$ equals to 0.25 or 0.29 for different attacks.

(S2) Simulate the first query. (I) Key generation. A ROK with respect to $N$ and $p$ is generated. More concretely, there are $N$ bits in this key and Alice knows every key bit with probability $p$. (II) Dilution. FOK is obtained from the above ROK according to the $N-N$ method. The state of every FOK bit, i.e. known, unknown, or almost known (equivalently, belonging to an almost known set), is determined after this step. (III) Record maintenance. Alice announces a shift $s$ $(0\leq s\leq N-1)$, which can be selected at random here, so that Bob encrypts the database via the shifted FOK and sends the whole ciphertext to her. Then Alice records the state of every item accordingly. Obviously, the items encrypted via known (unknown) key bits are still know (unknown), and the ones encrypted via an almost known key set still compose an AKS.

(S3) Simulate another query. The first two steps (I) and (II) are the same as that in (S2). (III) Record maintenance. Alice announce an optimal shift $s$ $(0\leq s\leq N-1)$ so that Bob encrypts the database via the shifted FOK and sends the whole ciphertext to her. Here by \textit{optimal} we mean that when the shift is chosen the unknown information about the database $H=n_u+n_{aks}$ is the lowest, where $n_u$ is the total number of unknown items, and $n_{aks}$ is the total number of AKS after Alice received the ciphertext encrypted via this FOK. Afterwards, Alice updates the state of every item accordingly.

(S4) Repeat (S3) until all the items in Bob's database are explicitly known for Alice.

\begin{table}[b]
\caption{\label{tab:1}%
The DQA for different $N$. Here $\overline{n}$ represents the expected amount of items Alice will get via one query.}
\begin{ruledtabular}
\begin{tabular}{llll}
\vspace{1mm}$N$ & $225$ & $1024$ & $10^4$\\
\vspace{1mm}$k$ & 3 & 4 & 6\\
\vspace{1mm}$\overline{n}=Np^k$ $(p=0.25)$ & 3.52 & 4.00 & 2.44\\
\vspace{1mm}$\overline{q}_d$ \hspace{9.35mm}$(p=0.25)$ & 18.6 & 30.4 & 53.4\\
\vspace{1mm}$\overline{n}=Np^k$ $(p=0.29)$ & 5.49 & 7.24 & 5.95\\
\vspace{1mm}$\overline{q}_d$ \hspace{9.35mm}$(p=0.29)$ & 15.4 & 23.3 & 40.0\\
\end{tabular}
\end{ruledtabular}
\end{table}

We perform simulations for three values of $N$, that is, 225, 1024, and $10^4$, and both $p=0.25$ and $p=0.29$ are executed for different $N$. For each situation the simulation is done over 10 runs and the average amount of queries (i.e. $\overline{q}_d$), which is needed for Alice to steal the whole database, is obtained and shown in Table \ref{tab:1}. Here we call $\overline{q}_d$ the Death Query Amount (DQA).
It can be seen that the leakage of parity information in $N-N$ method seriously damages the security of Bob's database. For example, when $N=10^4$ a dishonest Alice can steal the whole database after only 53.4 (for $p=0.25$) queries on average. While, as is shown, the expected amount of items Alice will get via one query is 2.44, and consequently the DQA should have been at least $10^4/2.44=4098.4$. Note that in this attack what Alice needs to do is only legally performing multiple queries (or equivalently, collecting data from other users who collude with her), and this insecurity comes from the leakage of parity information completely. If Alice executes a more complex attack, e.g. using individual USD measurement so that she can get any ROK bit with probability 0.29 instead of 0.25, the DQA would be further decreased to 40.0.

More concretely, three typical simulation instances for $N=10^4$, 1024, and 225 are shown in Figs. 1, 3, and 4, respectively. From these figures we can see how the amount of known or almost known items changes with respect to the query count $n_q$. Fig. 2 demonstrates how the theoretical amount of unknown information about the database changes in the instance $N=10^4$. Besides, one may also be interested in how many items Alice can explicitly obtain after each query. To show that we also depict the relation between the amount of explicitly known items and $n_q$. Obviously less queries (than DQA) might be enough if Alice wants to know only part of the database instead of the whole.

\begin{figure}
\centering
\subfigure[\hspace{0.03cm} $n_q$=1]{
\includegraphics[width=1in]{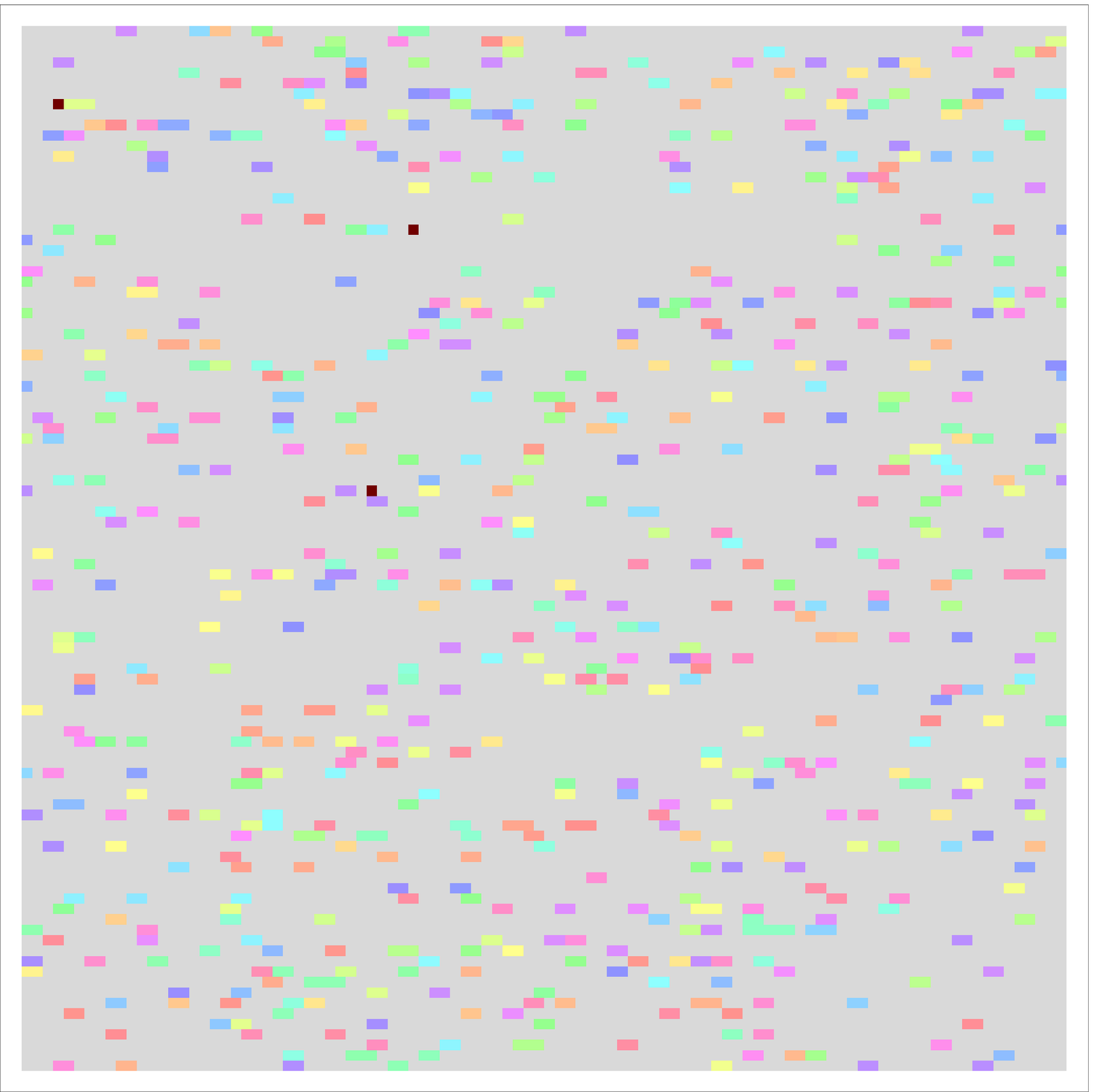}}
\subfigure[\hspace{0.03cm} $n_q$=7]{
\includegraphics[width=1in]{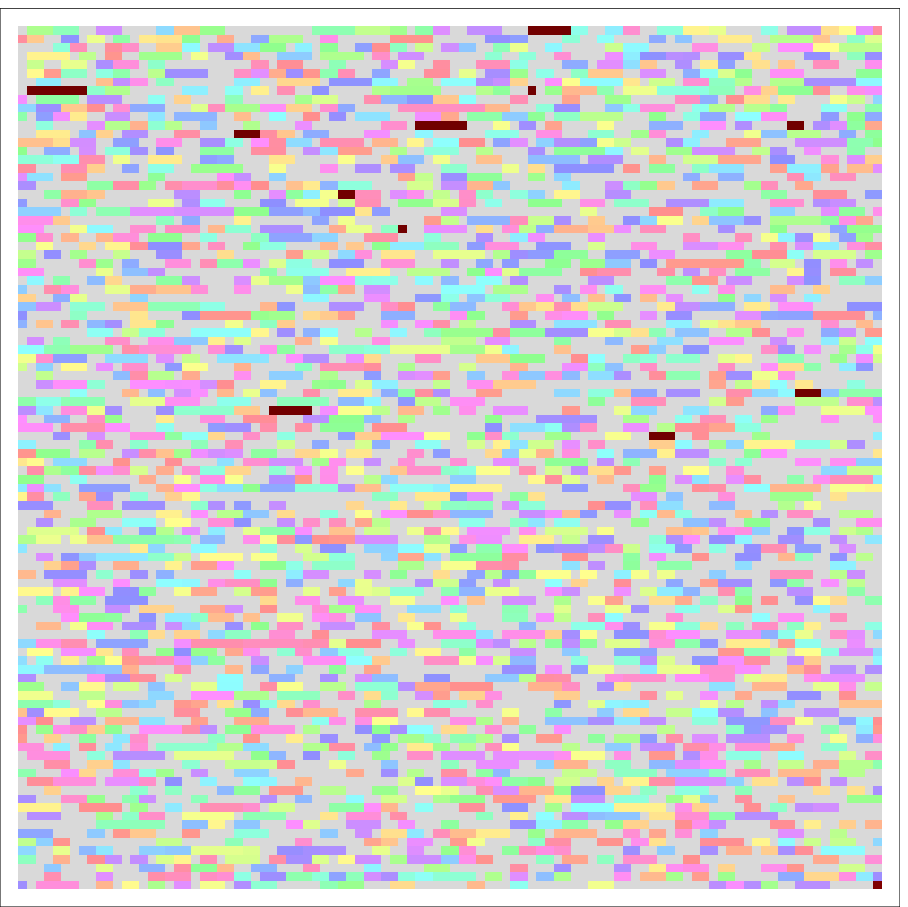}}
\subfigure[\hspace{0.03cm} $n_q$=14]{
\includegraphics[width=1in]{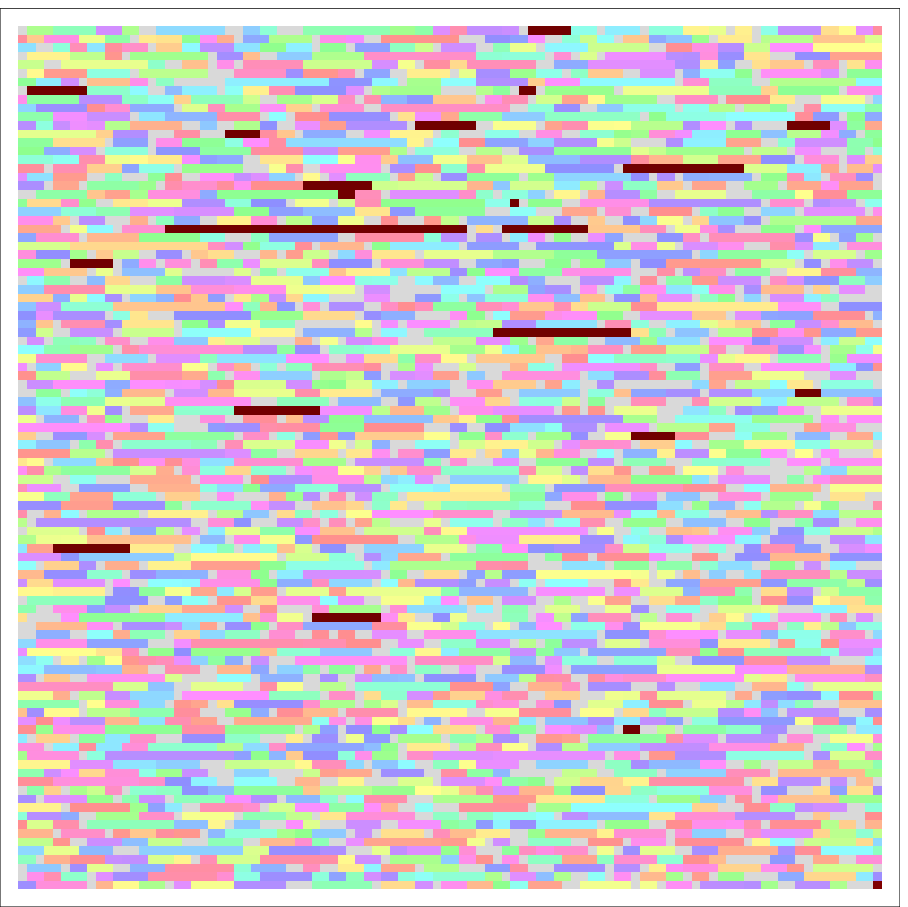}}\\
\subfigure[\hspace{0.03cm} $n_q$=21]{
\includegraphics[width=1in]{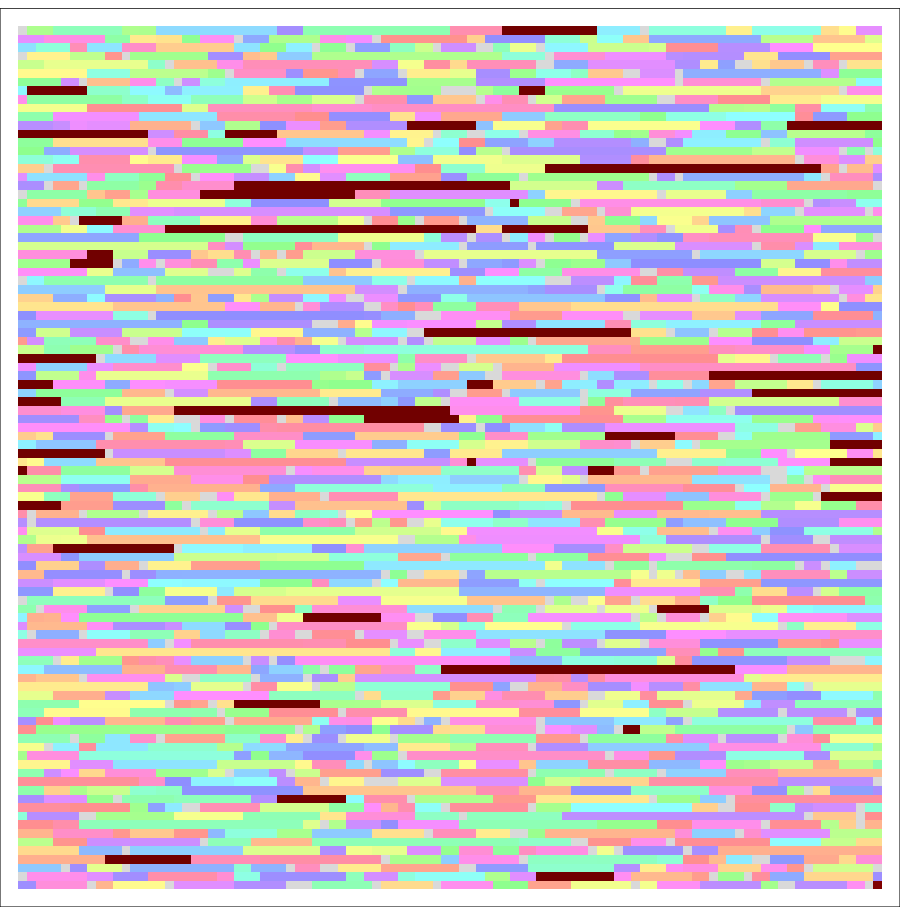}}
\subfigure[\hspace{0.03cm} $n_q$=28]{
\includegraphics[width=1in]{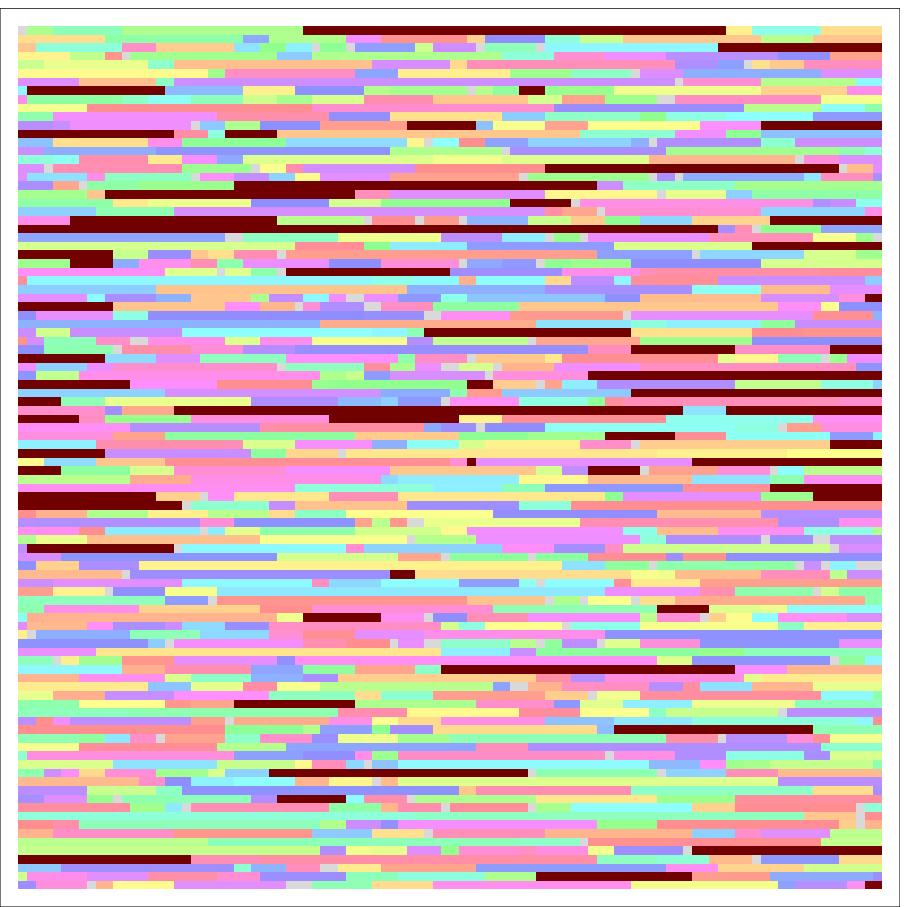}}
\subfigure[\hspace{0.03cm} $n_q$=35]{
\includegraphics[width=1in]{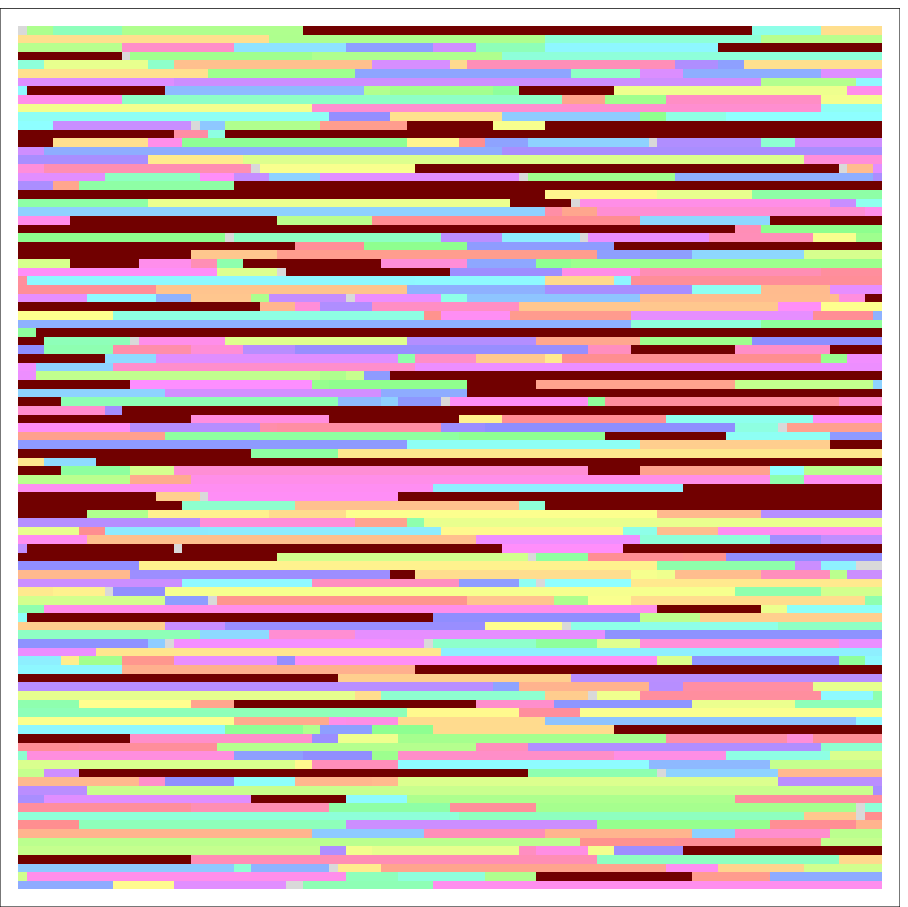}}\\
\subfigure[\hspace{0.03cm} $n_q$=42]{
\includegraphics[width=1in]{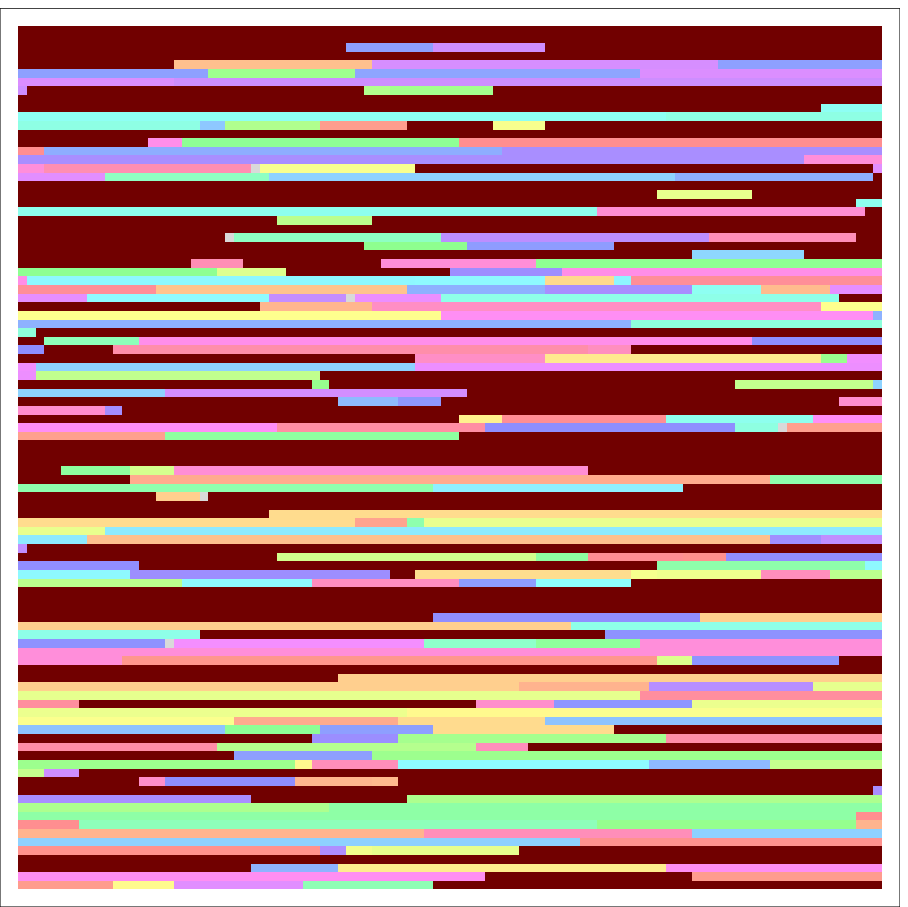}}
\subfigure[\hspace{0.03cm} $n_q$=49]{
\includegraphics[width=1in]{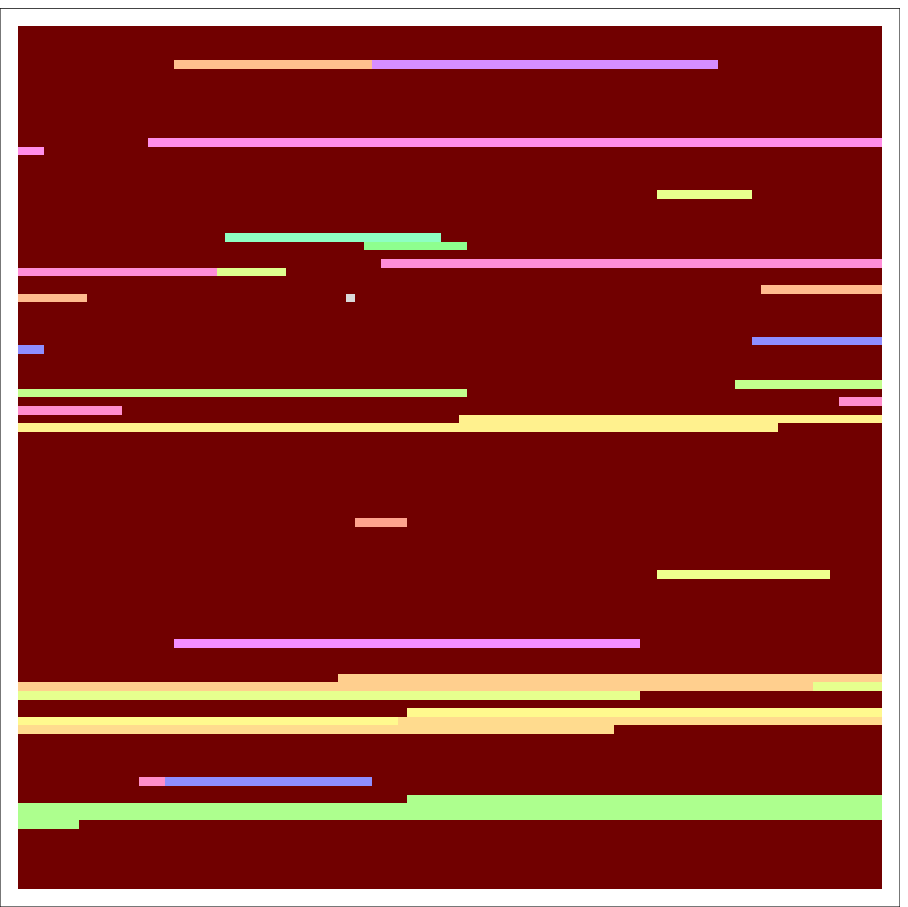}}
\subfigure[\hspace{0.03cm} $n_q$=53]{
\includegraphics[width=1in]{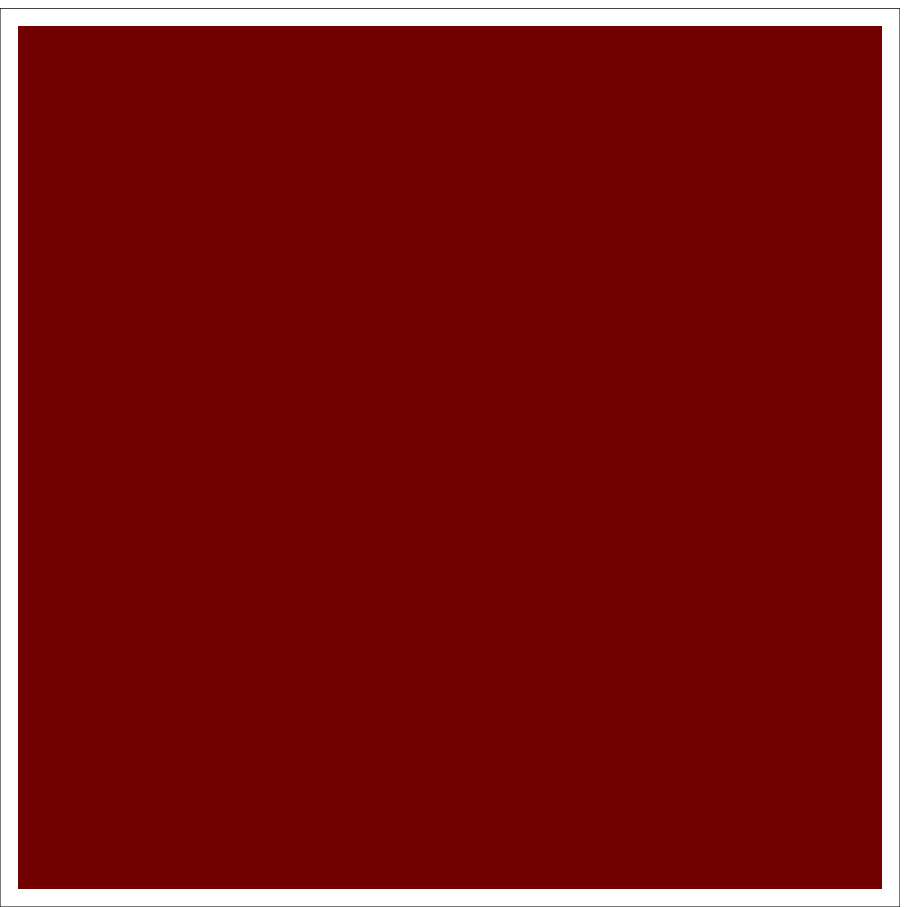}}\\
\hspace{1.2in}
\caption{A simulation instance for $N=10^4$, $k=6$, and $p=0.25$, which is directly adapted from Ref. \cite{PJ13}. To identify all the database items clearly, we draw a picture of the database, where any item is represented by a square, and dye the known items dark red, the unknown ones grey, and the AKS other different light colors. Here $n_q$ is the present count of queries and each sub-figure denotes the state of the whole database. In this instance the DQA is 53.
}
\label{fig:1} 

\end{figure}

\begin{figure}
\centering
\includegraphics[width=3in]{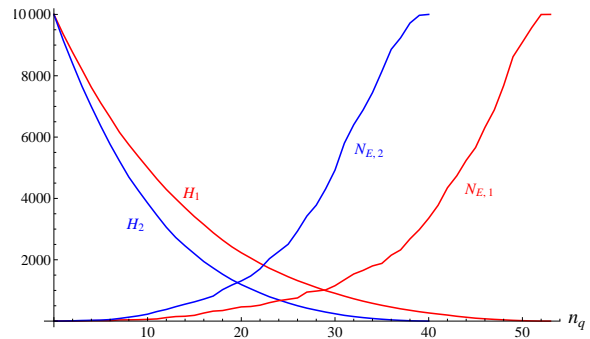}

\hspace{1in} \caption{The amount of unknown information about the database, denoted by $H$, and that of explicitly known items, denoted by $N_E$, after each query for $N=10^4$ and $k=6$. The red lines represent the result for an instance with $p=0.25$, i.e. the one shown in Fig. \ref{fig:1}, and the blue lines denote the result for an instance with $p=0.29$. To obtain the whole database, in the above two instances, Alice needs only 53 and 40 queries, respectively.}
\label{fig:2} 

\end{figure}

\begin{figure}
\centering
\subfigure[\hspace{0.03cm} $n_q$=1]{
\includegraphics[width=1in]{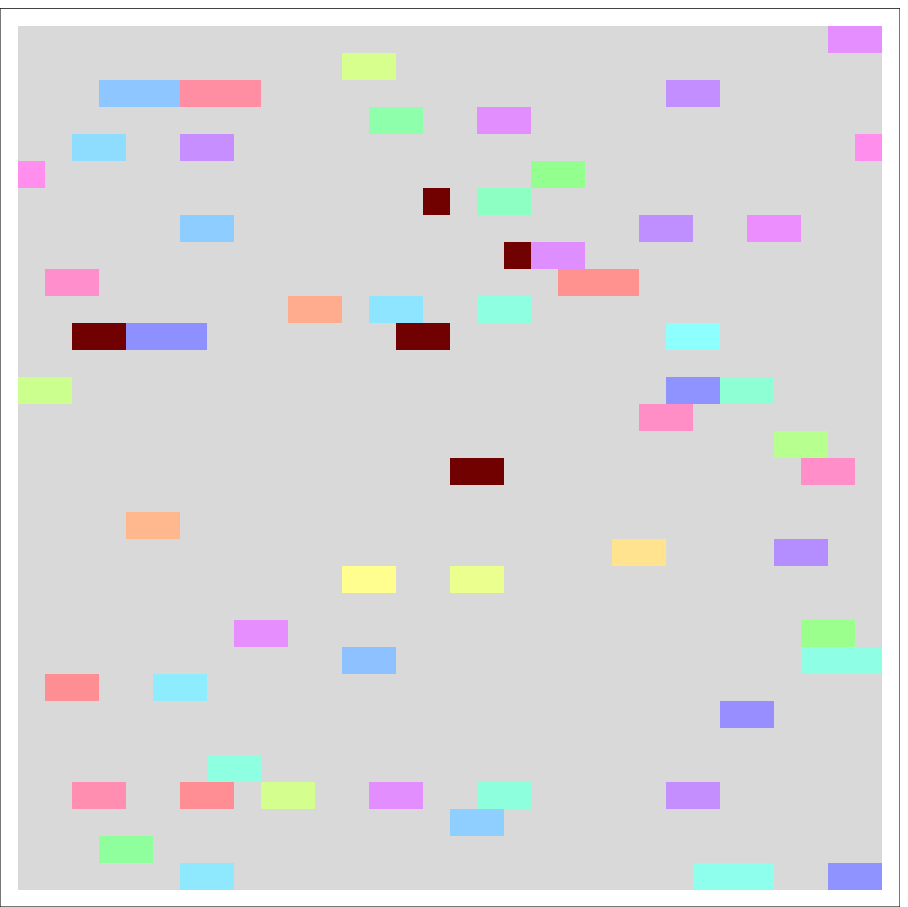}}
\subfigure[\hspace{0.03cm} $n_q$=7]{
\includegraphics[width=1in]{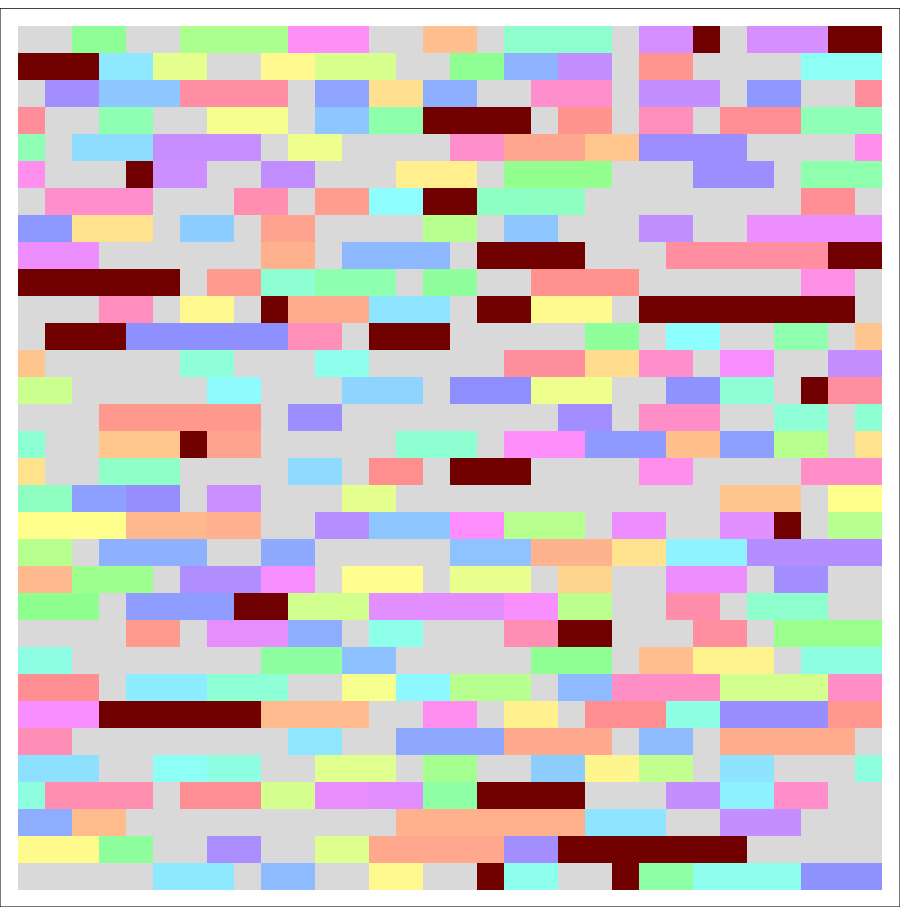}}
\subfigure[\hspace{0.03cm} $n_q$=13]{
\includegraphics[width=1in]{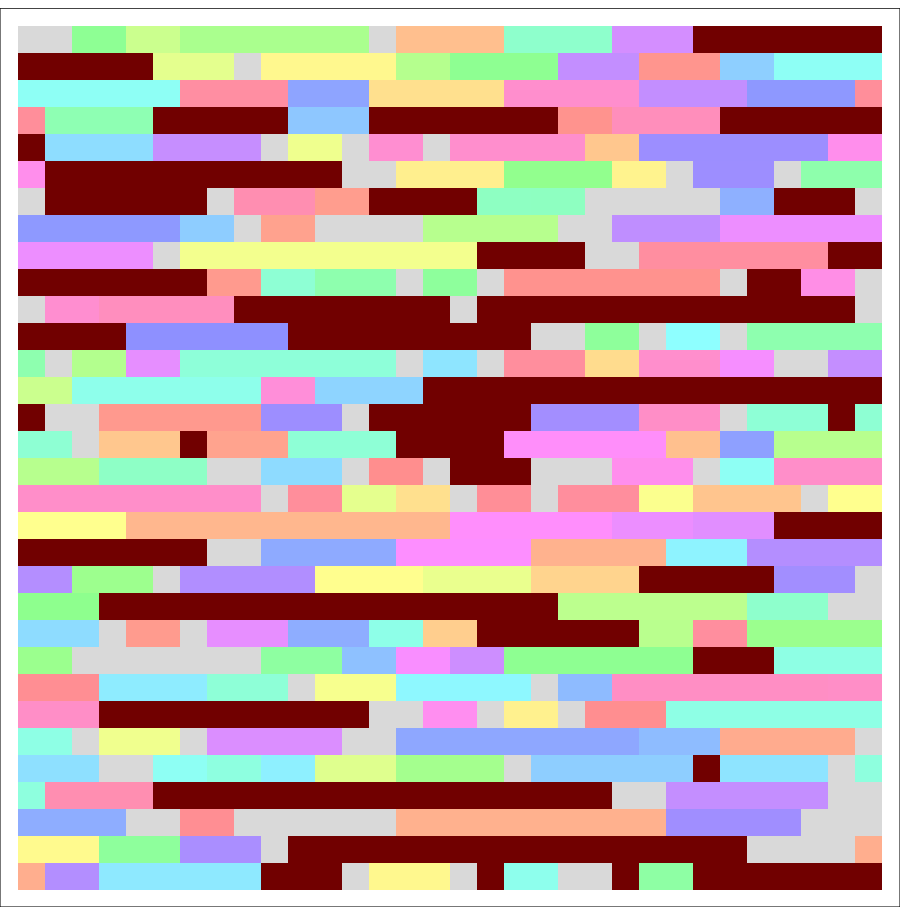}}\\
\subfigure[\hspace{0.03cm} $n_q$=19]{
\includegraphics[width=1in]{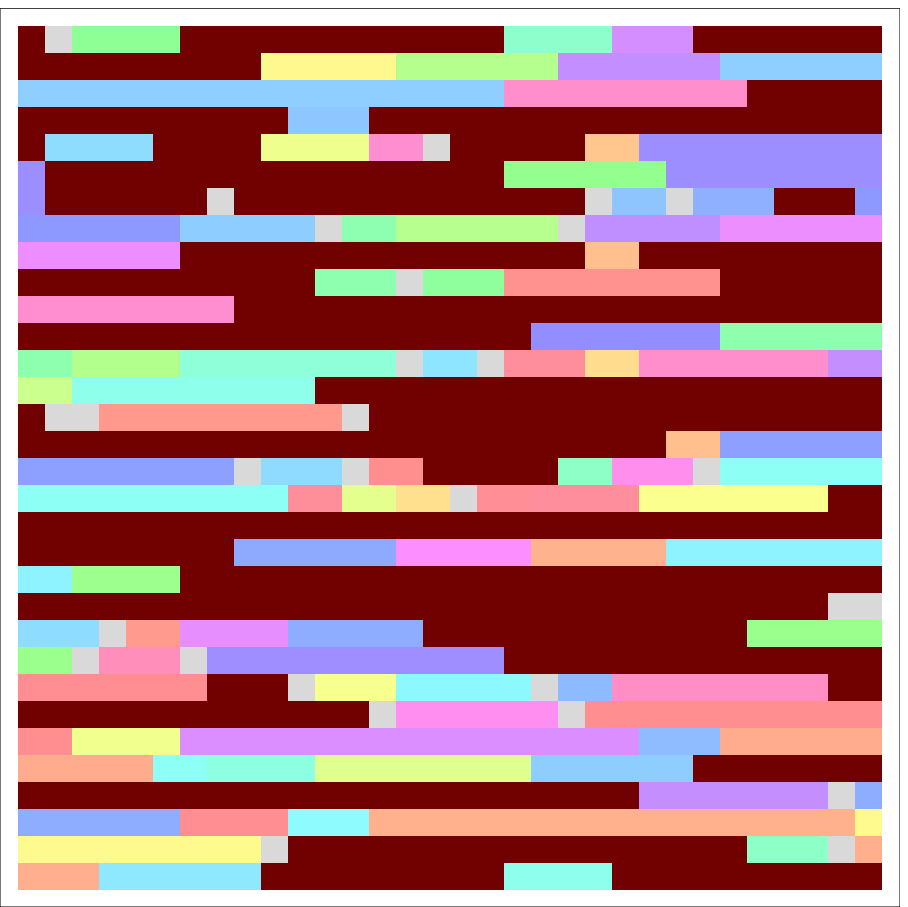}}
\subfigure[\hspace{0.03cm} $n_q$=25]{
\includegraphics[width=1in]{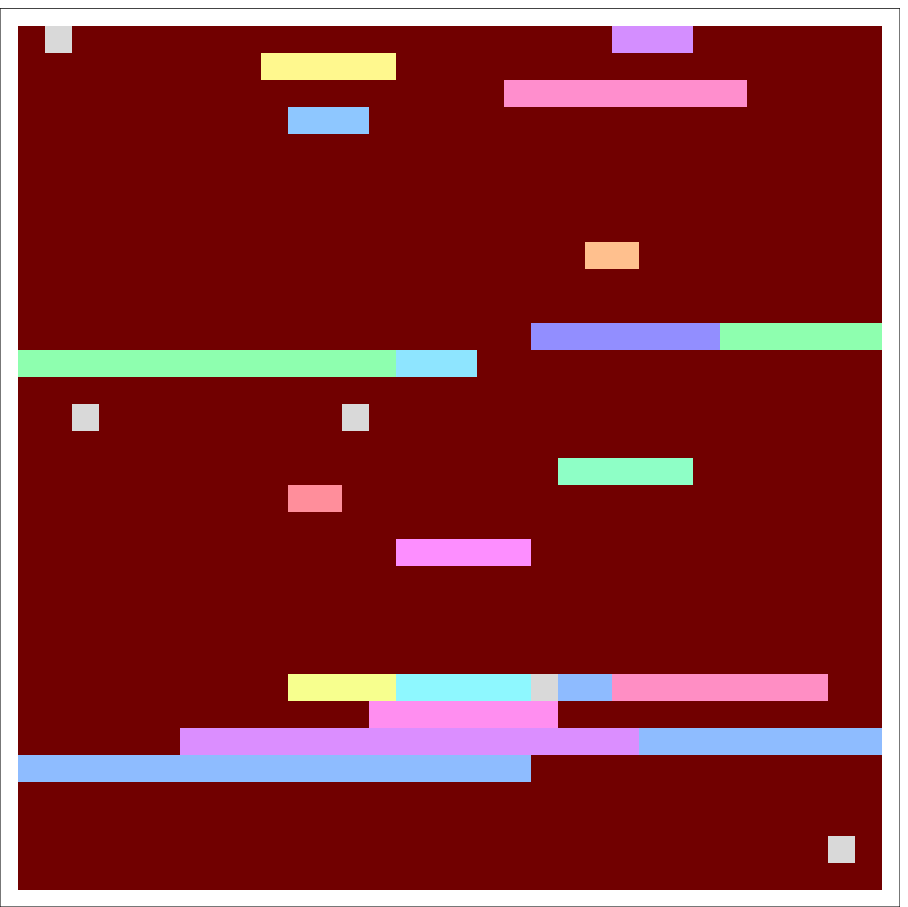}}
\subfigure[\hspace{0.03cm} $n_q$=30]{
\includegraphics[width=1in]{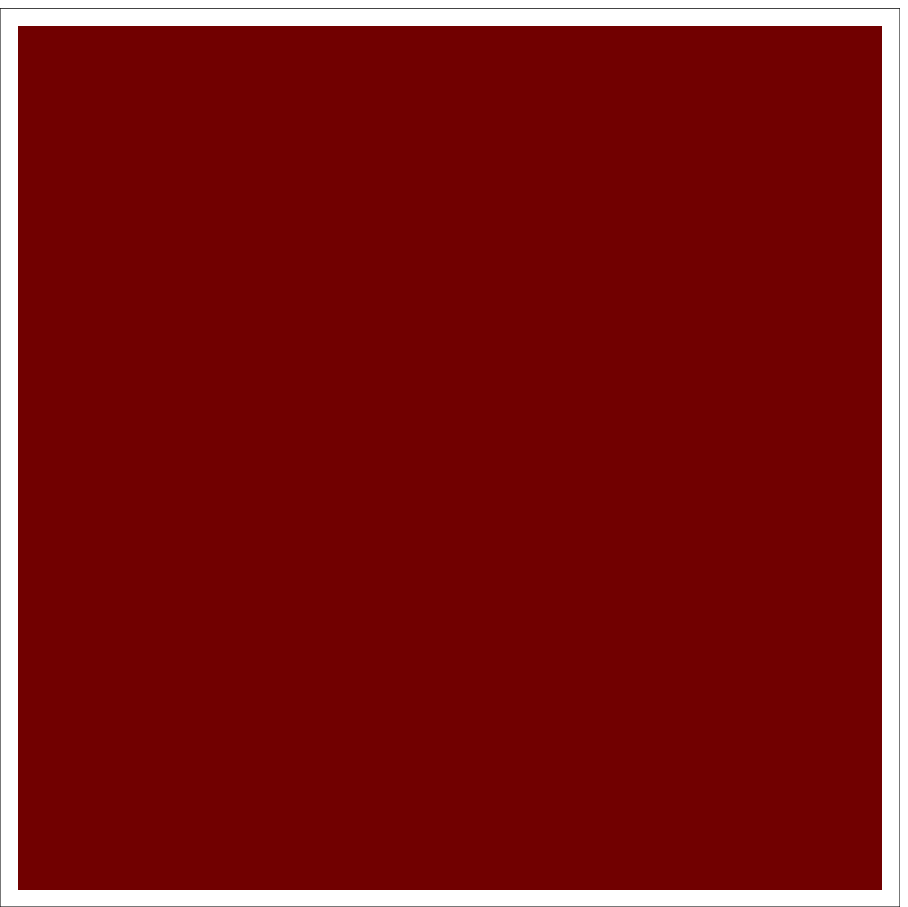}}\\
\hspace{1in}
\caption{A simulation instance for $N=1024$, $k=4$, and $p=0.25$. In this instance the DQA is 30.}

\label{fig:3} 
\end{figure}

\begin{figure}
\centering
\subfigure[\hspace{0.03cm} $n_q$=1]{
\includegraphics[width=1in]{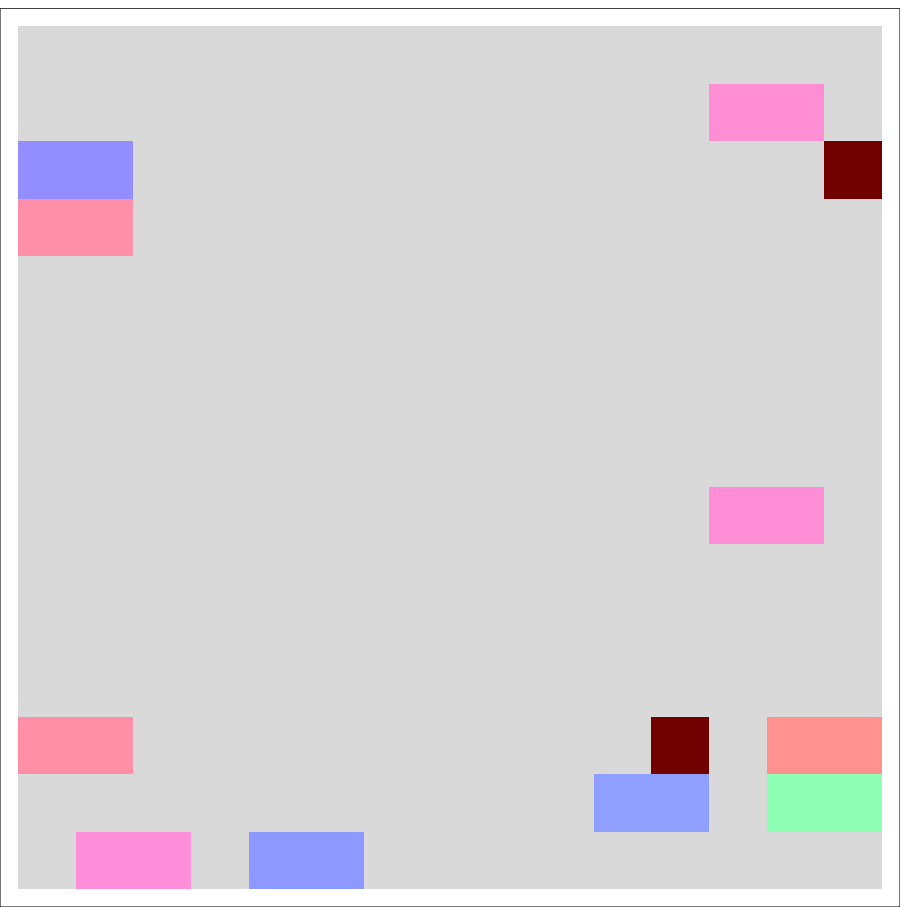}}
\subfigure[\hspace{0.03cm} $n_q$=5]{
\includegraphics[width=1in]{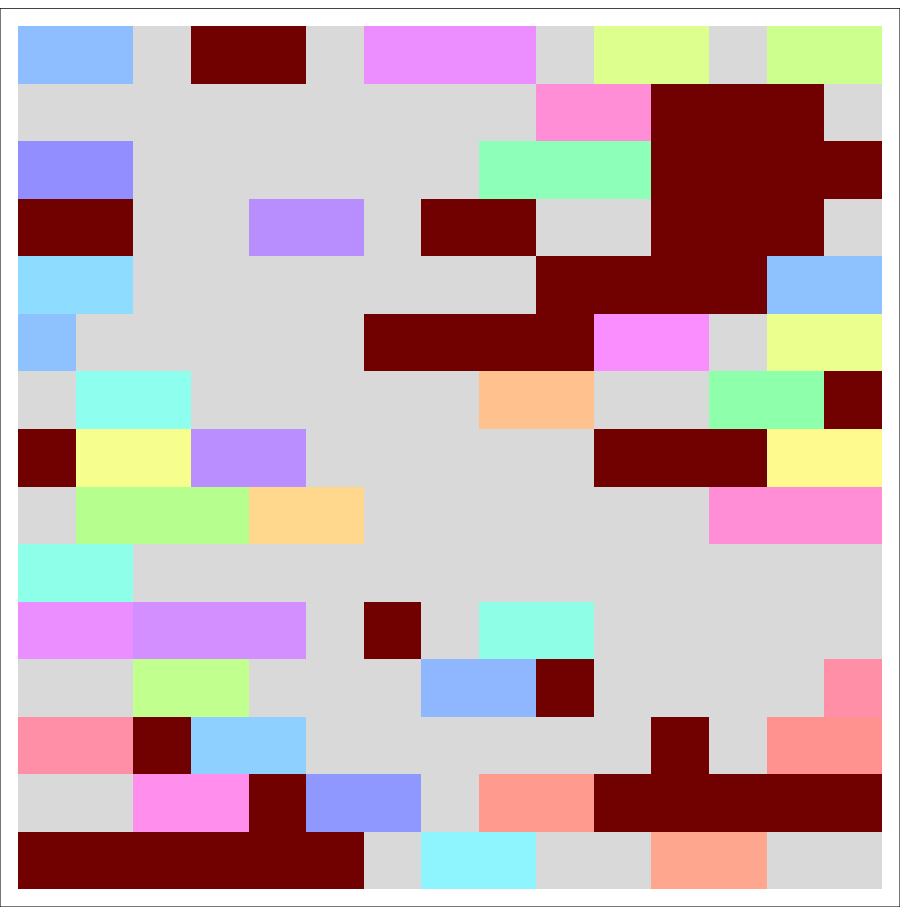}}
\subfigure[\hspace{0.03cm} $n_q$=9]{
\includegraphics[width=1in]{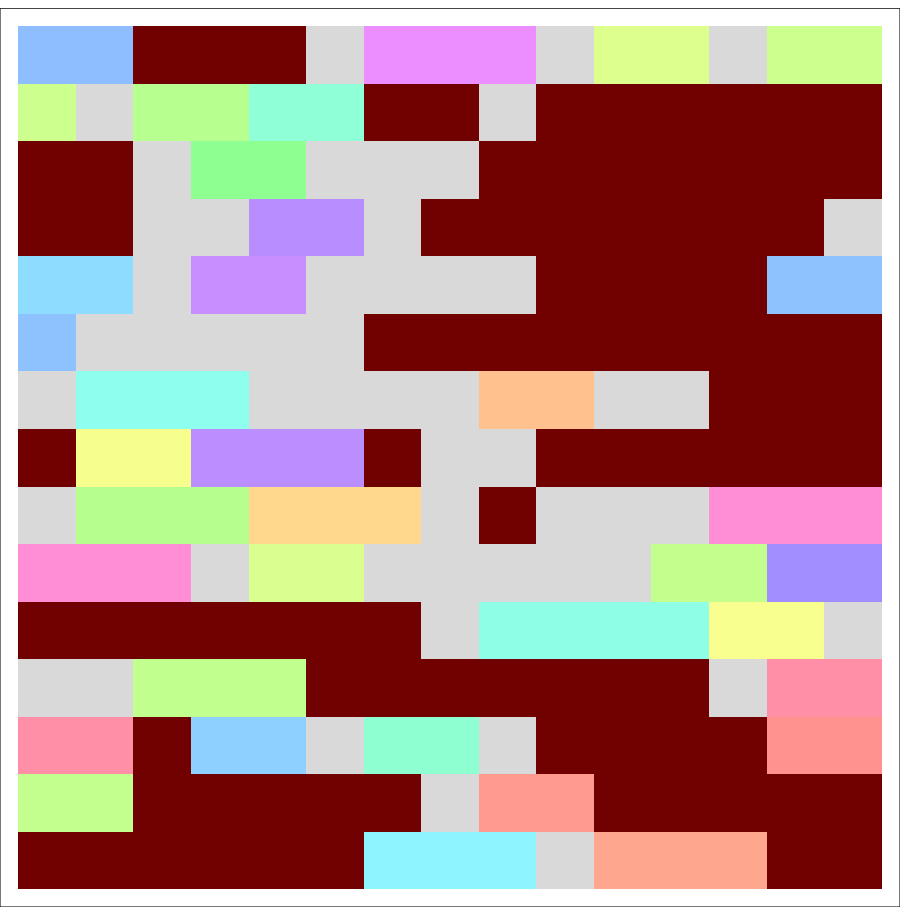}}\\
\subfigure[\hspace{0.03cm} $n_q$=13]{
\includegraphics[width=1in]{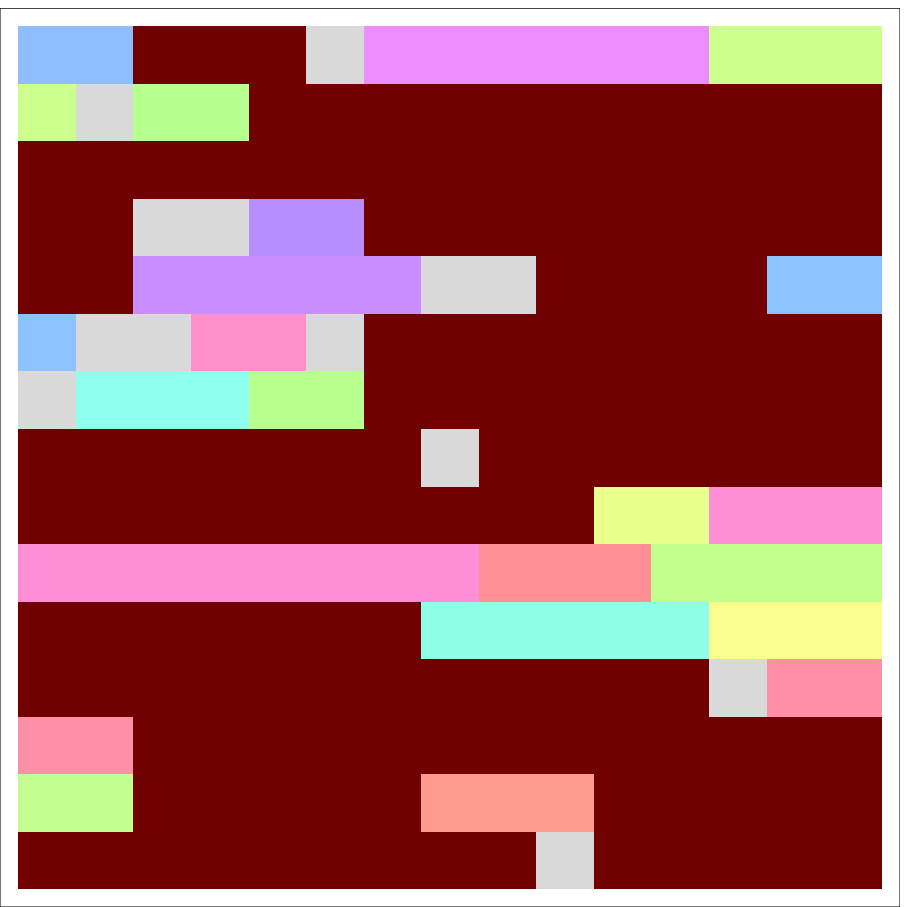}}
\subfigure[\hspace{0.03cm} $n_q$=17]{
\includegraphics[width=1in]{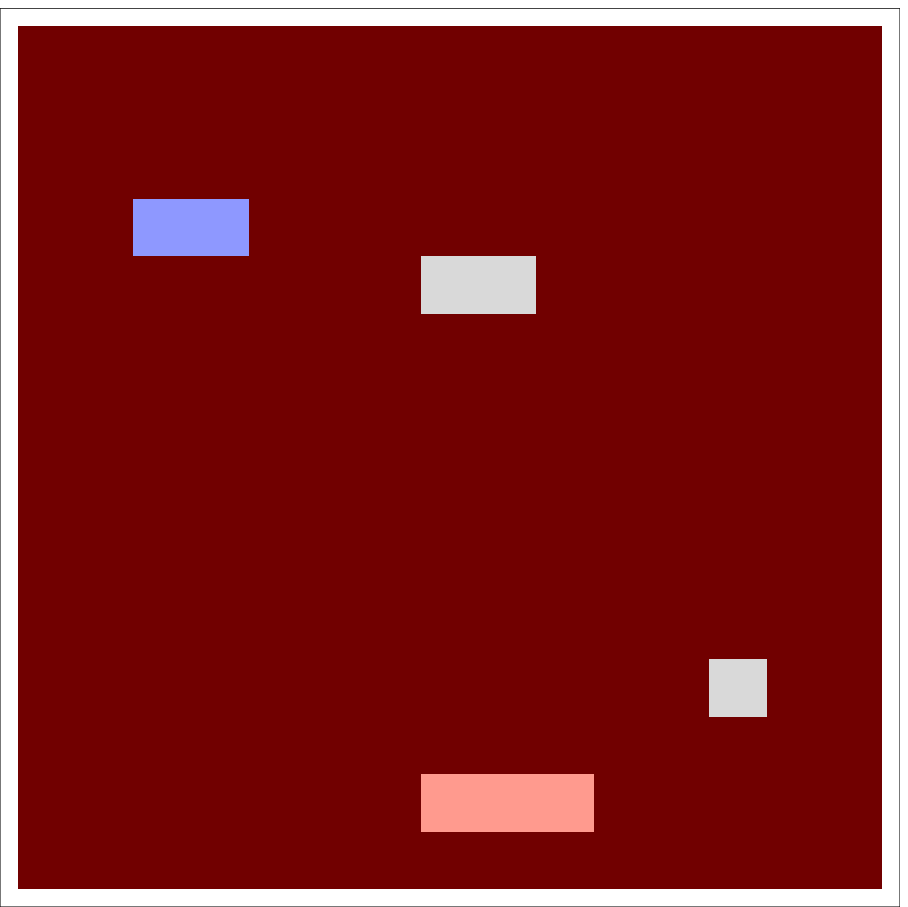}}
\subfigure[\hspace{0.03cm} $n_q$=19]{
\includegraphics[width=1in]{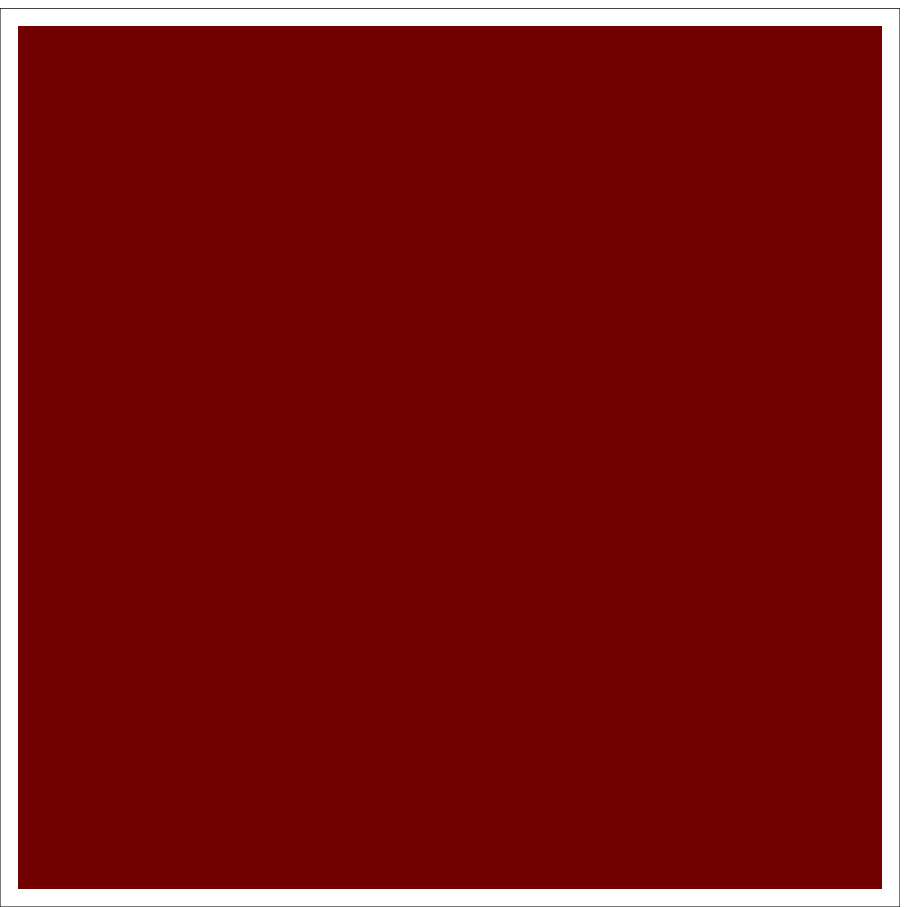}}\\
\hspace{1in}
\caption{A simulation instance for $N=225$, $k=3$, and $p=0.25$. In this instance the DQA is 19.}

\label{fig:4} 
\end{figure}

\subsection{Analysis on the $rM-N$ method}

In the $rM-N$ dilution method, Alice and Bob generate an $N$-bit FOK by $rM$ ROK bits. This method greatly reduces the communication complexity compared with the $kN-N$ method (generally $rM\ll N$). However, the information entropy of such an $N$-bit FOK is $rM$ at maximum, which means this FOK is not strictly secure in theory.
Through analysis, we get the following theorem.

\textbf{Theorem 1.}
In a QPQ protocol which employs the $rM-N$ dilution method, the user Alice can get the total database by at most $rM$ queries.

\begin{proof}
We start with the analysis on the structures of the FOK $O^F$ generated by $rM-N$ dilution method. If the sub-key $O^{R_i}$ (with length $M$), the extended sub-key $\widetilde{O}^{R_i}$ (with length $N$), and the FOK $O^F$ are seen as column vectors, we can write the dilution method as
\begin{equation}\label{ofg}
    O^F=\bigoplus\limits_{i=1}^r\widetilde{O}^{R_i}(s_i)=\bigoplus\limits_{i=1}^rG(s_i)O^{R_i},
\end{equation}
where $\widetilde{O}^{R_i}(s_i)$ represents the new key generated by performing a relative shift $s_i$ on $\widetilde{O}^{R_i}$, and $G(s_i)$ is an $N\times M$ matrix. Obviously here $G(s_i)$ includes both the sub-key-extension operation and the key-shift operation. According to the processes described in subsection \ref{rm-n}, we can get that the row vectors of $G(s_i)$ are $N$ different $M$-dimensional binary vectors, each contains $k$ 1s and $M-k$ 0s. Furthermore, different key-shift operations (that is, different $s_i$) just implies different orders of the row vectors in $G(s_i)$ \cite{fn}.

To give further analysis on the structures of $O^F$, we consider the equation
\begin{eqnarray}\label{ORK1}
    & &\begin{bmatrix}
    \widetilde{O}^{R_1}(s_1) & \widetilde{O}^{R_2}(s_2) & \cdots & \widetilde{O}^{R_r}(s_r)\end{bmatrix}\nonumber\\
    &=&\begin{bmatrix} G(s_1) & G(s_2) & \cdots & G(s_r) \end{bmatrix} \begin{bmatrix} O^{R_1} & & &\\ & O^{R_2} & &\\ & & \ddots &\\&&&O^{R_r} \end{bmatrix}.
\end{eqnarray}
For simplicity, here we denote the above three matrices as $\widetilde{O}_s$, $G_s$, and $O^R_I$ respectively, and Eq. (\ref{ORK1}) can be written in the form of $\widetilde{O}_s=G_s O^R_I$. By rewriting the first two matrices in the form of row vectors, Eq. (\ref{ORK1}) can be expressed as
\begin{eqnarray}\label{ORK2}
   \begin{bmatrix}
    o^1 \\ o^2 \\ \vdots \\o^N\end{bmatrix}
    =
    \begin{bmatrix} g^1 \\ g^2 \\ \vdots \\ g^N  \end{bmatrix} O^R_I
    =
    \begin{bmatrix} g^1O^R_I \\ g^2O^R_I \\ \vdots \\ g^NO^R_I  \end{bmatrix},
\end{eqnarray}
where $o^j$ and $g^j$ ($j\in [1,N]$) are row vectors. Suppose the rank of $G_s$ is $T$, then $T\leq rM$ since $G_s$ is an $N\times rM$ matrix. Suppose $\{g^{\gamma_1}, g^{\gamma_2}, \ldots, g^{\gamma_T}\}$ is a maximally linearly independent set of the row vectors of $G_s$, which means any row vector of $G_s$ can be expressed by these $T$ vectors, that is
\begin{eqnarray}
g^j=\bigoplus_{t=1}^{T}\lambda_{j,t}g^{\gamma_t}, \forall j\in [1,N],
\end{eqnarray}
where $\lambda_{j,t}=0,1$. Then for any $j\in [1,N]$ we have
\begin{eqnarray}\label{lrot}
 o^j=g^jO^R_I=(\bigoplus_{t=1}^{T}\lambda_{j,t}g^{\gamma_t})O^R_I=\bigoplus_{t=1}^{T}\lambda_{j,t}o^{\gamma_t},
\end{eqnarray}
and consequently we arrive at
\begin{eqnarray}\label{lrOF}
 O^F_j=\bigoplus\limits_{i=1}^{r}o^j_i=\bigoplus\limits_{i=1}^{r}\bigoplus_{t=1}^{T}\lambda_{j,t}o^{\gamma_t}_i=\bigoplus_{t=1}^{T}\lambda_{j,t}O^F_{\gamma_t}.
\end{eqnarray}
Eqs. (\ref{lrot}) and (\ref{lrOF}) imply that the $T$ vectors $\{o^{\gamma_t}|t\in [1,T]\}$ can linearly express all the row vectors in $\widetilde{O}_s$, and $O^F$ is totally determined by the $T$ bits $\{O^F_{\gamma_t}|t\in [1,T]\}$. Therefore, \textit{once Alice gets these $T$ bits, she can figure out the whole $O^F$, and consequently obtain the whole database by decrypting the ciphertext of the database encrypted by the key $O^F$}. Here we call such a set of $T$ $O^F$ bits, which can determine the whole $O^F$, a basis of $O^F$. For example, the above $\{O^F_{\gamma_t}|t\in [1,T]\}$ is a basis of $O^F$.

Now we introduce Alice's attack strategy where she can get the total database by at most $rM$ queries. Suppose Bob's database is bit string $D=D_1D_2\ldots D_N$, where each bit $D_j$ $(j\in[1,N])$ represents an item. The processes of the attack is as follows.
\begin{itemize}
  \item [1] Alice honestly executes the first query to Bob's database. Suppose the $N\times rM$ matrix, i.e. the second one in Eq. (\ref{ORK1}), is $G_s^1$, the shifted FOK is $O^{F_1}(s'_1)$, where $s'_1$ is the shift value Alice chose to the FOK $O^{F_1}$, and one of Alice's known bits in $O^{F_1}(s'_1)$ is $O^{F_1}_{\gamma_1}(s'_1)$ (i.e. the $\gamma_1$-th bit in it). In this query, Alice will receive a ciphertext of the database $C^1$, in which the $j$-th ($j\in [1,N]$) bit is $C^1_j=D_j\oplus O^{F_1}_j(s'_1)$.

       Afterwards, according to $G_s^1$ and $s'_1$, Alice calculates a basis of $O^{F_1}(s'_1)$ which contains $O^{F_1}_{\gamma_1}(s'_1)$, denoted as $\{O^{F_1}_{\gamma_t}(s'_1)|t\in [1,T]\}$.
  \item [$t$] ($t=2, 3, \ldots, T$) Alice executes the $t$-th query to Bob's database. Suppose the FOK in this round is $O^{F_t}$ and one of the bits known to Alice is the $\nu_t$-th one. Then she declares a shift value $\gamma_t-\nu_t$, thus she knows the $\gamma_t$-th bit in the shifted FOK $O^{F_t}(\gamma_t-\nu_t)$. Consequently, she can calculate $O^{F_1}_{\gamma_t}(s'_1)$ when she receives the ciphertext of the database $C^t$.The reason is as follows. The $\gamma_t$-th bit in the two ciphertexts $C^1$ and $C^t$ are
\begin{eqnarray}
    C^t_{\gamma_t}&=&D_{\gamma_t}\oplus O^{F_t}_{\gamma_t}(\gamma_t-\nu_t),\nonumber\\
    C^1_{\gamma_t}&=&D_{\gamma_t}\oplus O^{F_1}_{\gamma_t}(s'_1),
\end{eqnarray}
so,
\begin{eqnarray}
    O^{F_1}_{\gamma_t}(s'_1)=C^t_{\gamma_t}\oplus C^1_{\gamma_t}\oplus O^{F_t}_{\gamma_t}(\gamma_t-\nu_t).
\end{eqnarray}
        \item [$T$+1]According to the $T$ known bits $\{O^{F_1}_{\gamma_t}(s'_1)|t\in [1,T]\}$ and the linear relationships determined by $G_s^1$ and $s'_1$, Alice calculates all the other bits in $O^{F_1}(s'_1)$ and then the whole database $D=C^1\oplus O^{F_1}(s'_1)$.
\end{itemize}

In the above attack strategy, Alice gets the whole database by only $T$ queries, where $T\leq rM$. Actually, Alice always needs less than $T$ queries. On the one hand, Alice may know more than one FOK bit in each query. On the other hand, there are many different bases of $O^{F_1}(s'_1)$, and she can choose the basis which contains more known bits to reduce the total number of queries.
\end{proof}

\section{Post-processing of the oblivious key with error correction}

As we know, the major advantage of QKD-based QPQ protocols is its practicability. In a practical realization, however, noise exists in the channel and there are always errors in the shared (raw) key between Alice and Bob. Therefore, error correction is necessary for such protocols. In fact, potential errors will seriously damage the function of QPQ. An error bit in Alice's FOK implies that Alice would pay her money and get back a wrong message from Bob, which is obviously unfair for Alice. Moreover, Alice would think Bob is dishonest if she got a wrong item, and Bob might cover his attack by channel noise (not that Bob's attack will inevitably result in giving a wrong message to Alice with a certain probability \cite{QPQ11}, and he can excuse that it is a result of channel noise when such error happens). But till now, as an open question in QPQ listed by M. Jakobi et al \cite{QPQ11}, the method of error correction in QKD-based QPQ protocols is still missing, which greatly limits the practicability of them.

Actually, performing error correction on oblivious key is difficult. On the one hand, Alice only knows part of the oblivious key, which is quite different from that of general communication or QKD. On the other hand, to correct errors, additional two-way communication is generally needed, which might affect the privacy of both users. Furthermore, this kind of influence is difficult to analyze \cite{QPQ11}. Here, enlightened by the idea of one-way error correction in QKD \cite{proof}, we present an error correction manner for the oblivious key. Together with previous $kN-N$ dilution and the technique of shift-addition, we actually give a \textit{complete} post-procession algorithm for the oblivious key in QPQ, including both dilution and error correction. Furthermore, the influence of error correction on the user's privacy is analyzed.

As analyzed above, the improved dilution method, i.e. $N-N$ and $rM-N$ ones, will result in the insecurity of Bob's database. So, let us go back to the original $kN-N$ method \cite{QPQ11} to reduce Alice's known bits, and take the scenario where $N=10^5$, $k=7$, $p=0.25$, $\overline{n}=6.10$ and the failure probability $P_0=(1-p^k)^N=0.002$ as our example, which is directly adapted from Ref. \cite{QPQ11}. In this condition Alice and Bob share a ROK $O^R=O^R_1O^R_2...O^R_{kN}$, with length $kN$, in the sense that Bob knows all the key while Alice knows every bit with probability $p=0.25$. After the dilution, a FOK $O^F=O^F_1O^F_2...O^F_N$, with length $N$, can be obtained according to Eq.(\ref{eq:one}).

Consider one FOK bit $O^F_i$, which equals to the parity of 7 ROK bits $\{O^R_{i+jN} (j=0,1,...,6)\}$. Suppose the error rate of Alice's ROK is $e$, that is, owing to the noise in quantum channel, Alice's every known bit in ROK differs from Bob's with probability $e$ (it is reasonable to assume that the error probabilities of $\{O^R_{i+jN}\}$ are independent because they come from different photons which are distant from each other). Similar to that in QKD, Alice and Bob can estimate this error rate by publicly comparing their key bits. For example, Bob announces part of his ROK bits, Alice compares them with the corresponding known ones in her ROK, and then declares the error rate.

Then, if there is no manner to correct errors, the error probability of the FOK bit $O^F_i$ is
\begin{eqnarray}
p_e=\sum_{t=1,3,5,7}(^7_t)e^t(1-e)^{7-t},
\end{eqnarray}
that is, the probability that odd errors happen in $\{O^R_{i+jN}\}$.

In fact, the users can correct errors in ROK and obtain FOK by the following steps. Note that here, for simplicity, we still take the above example to describe our method. That is, we only discuss how to correct errors and get a FOK bit from 7 ROK bits $\{O^R_{i+jN}\}$. Other ROK bits can be managed by the same manner.

(C1) Bob generates a message $V$ containing 4 random bits, i.e. $V=V_1V_2V_3V_4$, and encodes it into a 7-bit codeword $CW$ via a [7,4,3] linear error-correction code. For example, the code has a generator matrix
\begin{eqnarray}
G=\left[ \begin{array}{c c c c c c c} 1 & 0 & 0 & 0 & 1 & 0 & 1\\
0 & 1 & 0 & 0 & 1 & 1 & 1\\
0 & 0 & 1 & 0 & 1 & 1 & 0\\
0 & 0 & 0 & 1 & 0 & 1 & 1
\end{array} \right],
\end{eqnarray}
and it encodes 0000, 0001, 0010, ..., 1111 into
\begin{eqnarray}
0000000, \quad 0001011, \quad 0010110, \quad 0011101\nonumber\\
0100111, \quad 0101100, \quad 0110001, \quad 0111010\nonumber\\
1000101, \quad 1001110, \quad 1010011, \quad 1011000\nonumber\\
1100010, \quad 1101001, \quad 1110100, \quad 1111111
\end{eqnarray}
respectively. This code has a minimum distance $d=3$ and can correct 1 error among the 7 bits.

(C2) Bob encrypts the above 7-bit codeword $CW$, using $\{O^R_{i+jN}\}$ as the key, via one-time pad, and sends the ciphertext $c$ to Alice.

(C3) This step can be divided into the following two different conditions.

\textbf{Condt.1} Alice knows all the 7 ROK bits $\{O^R_{i+jN}\}$. In this condition, (I) Alice decrypts $c$, getting a 7-bit codeword $CW'$. (II) Alice corrects the possible error in $CW'$, obtaining $CW$. Obviously, if there are no more than 1 bit error in $\{O^R_{i+jN}\}$ Alice will obtain $CW$ correctly. (III) Alice takes the parity of 7 bits in $CW$ as the FOK bit $O^F_i$.  

\textbf{Condt.2} Alice does not know all the 7 ROK bits $\{O^R_{i+jN}\}$. In this condition, Alice labels this FOK bit unknown, i.e. $O^F_i=?$.

(C4) Bob also calculates the parity of 7 bits in $CW$ to obtain his corresponding FOK bit $O^F_i$.

By the above manner, Alice and Bob will finish their dilution and error correction of the ROK after the management of all the $7N$ ROK bits, obtaining $N$-bit FOK $O^F$. It is not difficult to see that, by introducing the error correction method, the error rate in FOK now becomes
\begin{eqnarray}
p'_e=\sum_{t=3,5,7}(^7_t)e^t(1-e)^{7-t},
\end{eqnarray}
that is, the probability that 3, 5, or 7 errors happen in $\{O^R_{i+jN}\}$. Obviously this error rate is greatly lower than the one without error correction, i.e. $p_e$, when $e$ is small. And the failure probability, which implies Alice does not obtain any FOK bit, remains unchanged, that is
\begin{eqnarray}
P'_0=P_0=(1-p^k)^N=0.002.
\end{eqnarray}

Now we analyze the influence on users' privacy the above error correction brings.

\textbf{Bob's privacy.} In the above example, to achieve the function of error correction (i.e. to reduce the error rate of the FOK bit $O^F_i$), Alice has to know all 7 ROK bits $\{O^R_{i+jN}\}$, which happens with probability $p_1=p^7$. Owing to the utilization of the above [7,4,3] code, however, Alice can deduce $CW$ and then get $O^F_i$ as long as she knows no less than 4 bits among the 7 correctly. Therefore, if Alice is dishonest and gives up the function of error correction she will know every FOK bit with probability
\begin{eqnarray}
p_2=\sum_{t=4,5,6,7}(^7_t)p^t(1-p)^{7-t}.
\end{eqnarray}

Obviously, there is a huge gap between $p_1$ and $p_2$, which greatly influences the database security. More concretely, considering $N=10^5$ and $p=0.25$, the expected amount of FOK bits known by an honest Alice equals $\overline{n}_1=Np_1=6.10$, while that by a dishonest Alice is $\overline{n}_2=Np_2=7055.66$. That is to say, if a dishonest Alice gives up the function of error correction, where the error rate of her known FOK bits will be relatively higher, she will obtain much more items in the database than expected. An intuitionistic way to strengthen Bob's privacy is choosing a greater $k$ ($k=7$ in the above example) to reduce the value of $\overline{n}_2$. But it is useless because it will greatly increase the failure probability $P_0$ (i.e. that of $\overline{n}_1=0$) for an honest Alice.

Now we must resolve this problem, that is, how to eliminate the gap between $p_1$ and $p_2$, though it is quite troublesome. Otherwise our above method for dilution and error correction will be useless. Fortunately, there is a manner, i.e. the technique of shift-addition (see subsection \ref{rm-n}), to overcome this difficulty. Recall that shift-addition has a good feature, that is, it can reduce Alice's known bits in the final key, while, at the same time, dose not increase the failure probability. Therefore, we can adapt this technique to deal with the oblivious key further. In detail,

(D1) Alice and Bob share $g$ ROK with length $7N$ via QKD protocol such as SARG04. Here $g$ is a parameter and its value will be discussed later.

(D2) Alice and Bob perform dilution and error correction on every ROK respectively, obtaining $g$ $N$-length middle oblivious key (MOK) $O^{M_i} \{i=1,2,...,g\}$ as they get FOK in steps (C1-C4).

(D3) To obtain the FOK $O^F$, the above $g$ MOKs are combined bitwise with relative shifts $s_i$ Alice can freely choose, that is
\begin{eqnarray}
O^F_j=\bigoplus_{i=1}^g O^{M_i}_{j+s_i}, 1\leq j\leq N,
\end{eqnarray}
where $O^F_j$ represents the $j$-th bit in $O^F$, and $O^{M_i}_{j+s_i}$ is the $(j+s_i)$-th bit in $O^{M_i}$.

Actually, the technique of shift-addition decreases the amount of FOK bits obtained by a dishonest Alice through distributing multiple oblivious keys. It really can resolve the above involved problem and render our error correct method to take effect though the communication efficiency becomes lower.

Obviously, our post-processing is a $gkN-N$ method with error correction. After this procedure, the final error rate in FOK $O^F$ is
\begin{eqnarray}
p''_e=\sum_{t=1,3,5..., t\leq g}(^g_t)(p'_e)^t(1-p'_e)^{g-t},
\end{eqnarray}
that is, the probability that odd errors happen in the $g$ MOK bits (which will be combined into one FOK bit). And the final failure probability is
\begin{eqnarray}
P''_0=1-(1-P'_0)^g,
\end{eqnarray}
which means that at least one MOK is totally unknown for Alice.

\begin{table}[b]
\caption{\label{tab:g}%
Simulation result for determining $g$ ($N=10^5$, $k=7$, $p=0.25$). Here $n_A$ represents the amount of FOK bits known by a dishonest Alice. Because $n_A$ remains unchanged from $g$=12 to $g$=20, we do not continue the simulation for $g>20$. In this simulation, with the value of $g$ increasing, $n_A$ would decrease rapidly to around 5 though the dishonest Alice has chosen an optimal shift $s_i$ for every MOK to make more bits survived.}
\begin{ruledtabular}
\begin{tabular}{lllllllll}
\vspace{1mm}$g$ & 1 & 2 & 3 & 4 & 5 & 6\~{}7 & 8\~{}11 & 12\~{}20\\
\vspace{1mm}$n_A$ & 7066 & 578 & 65 & 18 & 7 & 5 & 4 & 3
\end{tabular}
\end{ruledtabular}
\end{table}

Now we discuss how to choose the value of $g$. In fact we can perform simulations to determine $g$ and the amount of FOK bits known to a dishonest Alice, $n_A$. For the example here, i.e. $N=10^5$, our simulation result is given in Tab.\ref{tab:g}. It can be seen that $g=6$ is reasonable. In this condition $n_A=5$, which is already less than the expected amount of known FOK bits for a honest Alice in the original $kN-N$ method, i.e. $\bar{n}=6.10$. It should be emphasized that greater $g$ will bring less $n_A$, which implies higher database security, but it also results in lower communication efficiency.

\begin{figure}
\centering
\includegraphics[width=2.5in]{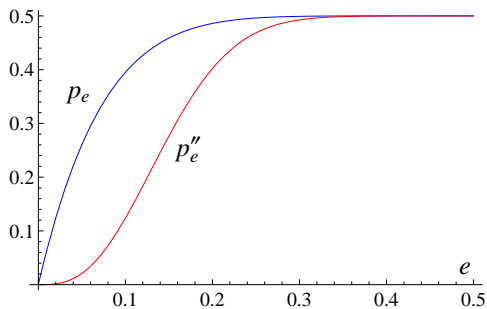}
\hspace{1in} \caption{The error rate of FOK. Here the line labeled by $p_e$ is the error rate in $kN-N$ method, and the other line $p''_e$ is that in our $gkN-N$ method.}
\label{fig:erate} 
\end{figure}

Fig. \ref{fig:erate} depicts the comparison between the error rates of FOK in the original $kN-N$ method and in our $gkN-N$ one. It can be seen that when $e<30\%$ the error rate of FOK in our $gkN-N$ method is visibly lower than that in the original $kN-N$ one, which implies the function of error correct takes effect in our method. For example, suppose $e=3\%$. An overall comparison between the original $kN-N$ method and our $gkN-N$ one is given in Tab. \ref{tab:overall}. Obviously, by a tolerable sacrifice on the failure probability (and communication efficiency), our method can significantly decrease the error rate of the final oblivious key.

\begin{table}[b]
\caption{\label{tab:overall}%
An overall comparison between the $kN-N$ dilution method and our $gkN-N$ post-processing for the example where $N=10^5$, $k=7$, $p=0.25$, $g=6$, and $e=3\%$. Here the 2nd column represents the expected amount of FOK bits which can be obtained by an honest Alice, and the 3rd one is that for a dishonest Alice (note that by ``dishonest'' we mean Alice tries to obtain more key bits by just giving up the function of error correction). The 4th column denotes the error rate in the final oblivious key, and the 5th one is the failure probability.}
\begin{ruledtabular}
\begin{tabular}{lllll}
\vspace{1mm} & $\bar{n}$-hon. & $\bar{n}$-dishon. & error rate & failure prob.\\
\vspace{1mm}$kN-N$ & 6.10 & - & 0.1758 & 0.002\\
\vspace{1mm}$gkN-N$ & 1 & 5 & 0.0008 & 0.013
\end{tabular}
\end{ruledtabular}
\end{table}

\textbf{Alice's privacy.} We believe that Alice's privacy in our $gkN-N$ method is the same as the $kN-N$ one. The reasons are as follows. (1) Alice's privacy in MOK in our method is the same as that in FOK in $kN-N$ one because there is no difference for Bob to gather information on a bit's conclusiveness in the above two keys (note that our error correction is executed in a one-way-communication style). That is, knowing a conclusive bit in Alice's MOK implies no information on this bit's value for Bob \cite{QPQ11}. (2) Suppose finally Alice knows the $i$th bit in FOK in our method. Because FOK is the combination of shifted MOKs, to know Alice's privacy, i.e. the index $i$, Bob has to know a conclusive bit in at least one of the MOKs. But if he obtains that Bob will totally loss the knowledge of this bit's value, and consequently not know the value of the $i$th bit in FOK. This security for Alice's privacy is just the same as that in $kN-N$ method.

Finally we emphasize the following two points about our post-processing, especially the error correction in it.

(1) Note that here, for simplicity, we use the parity of codeword as the FOK bit. To ensure the randomness of FOK, the code must has balanced parity, that is, half codewords have odd parity and the other half have even one. Of course we can also choose a code which has not balanced parity. In this condition, the parity of message (that is, the four-bit one corresponding to every codeword in our above example, e.g. 0000, 00001, 0010, ..., 1111) instead of codeword can be used as the FOK bit. Obviously the parity of message is naturally random.

(2) In the description of our post-processing, a [7,4,3] linear error-correction code is taken as our example. Of course we can also choose other code according to the requirement of error correction. For example, a code with a minimum distance d=5 is needed if we want correct any 1-bit and 2-bit error in the $k$ ROK bits whose parity will be a FOK bit. In our example, $k$ happens to be 7, and so the [7,4,3] code, in which the length of codewords is 7, is suitable. One may argue that our error correction may not work regularly if we want to use [7,4,3] code in some scenario where $k$ is not 7 (for example, as shown in Ref.~\cite{QPQ11}, when $N=10^4$ a reasonable value of $k$ is 6). In fact it is unnecessary to worry about it. According to the manner given in Ref.~\cite{OE12}, any expected value of $k$ can be achieved for different $N$ by adjusting the parameter $\theta$. Therefore, when we choose the code we only need to consider our requirement of error correction, and our post-processing method is universal for different scenarios. Of course, for the scenarios other than our above example (i.e. $N=10^5$, $k=7$ and $p=0.25$), the performances of our post-processing, including the error rates before and after the procedure, the user's privacy, the failure probability, and the value of $g$,  should be re-analyzed by similar manners.

\section{Conclusions}

From the proposing of BB84 QKD protocol \cite{BB84}, quantum cryptography has drawn much attention of the scholars in the world. Because of its success in the high security in key distribution, people hope the security of various kinds of protocols in classical cryptography can be overall upgraded by quantum manners. To this aim, different kinds of quantum protocols have been proposed, including quantum secret sharing, quantum secure direct communication, quantum digital signature, quantum coin flipping, quantum bit commitment, and so on. However, it seems that QKD is the most practical one till now. Other quantum protocols tend to have different shortages, e.g., failure in pursuing perfect security, excessive complexity, fragility against channel noise, or difficulty in realization. Therefore, in our opinion, it is valuable to study what kind of cryptographic aims can be achieved via QKD. It helps us to understand what kind of innovation quantum mechanics can bring to cryptography on earth. QKD-based QPQ is a good example for that, and it may push this study forward.

Though different protocols have been given, the post-processing of QKD-based QPQ is still unclear, which greatly limits the practicability of those protocols. Here we study the post-processing of the oblivious key in QKD-based QPQ, including both parts in it, i.e. dilution and error correction.

On the one hand, we demonstrate that, though they can significantly reduce communication complexity, the previous $N-N$ and $rM-N$ dilution methods \cite{PJ13} will result in insecurity in the sense that by multiple queries Alice can obtain much more items (even the whole database) than expected. For the $N-N$ method our simulation shows that when $N=10^4$ the dishonest Alice can steal the whole database after only 53.4 (for $p=0.25$) queries on average. While this number, called the Death Query Amount (DQA), should have been at least $N/\bar{n}=10^4/2.44=4098.4$. If Alice executes a more complex attack, e.g. using individual USD measurement, the DQA would be further decreased to 40.0. For the $rM-N$ method we prove that Alice can get the total database by at most $rM$ queries.

On the other hand, we propose an effective error-correction scheme for the oblivious key. Combined with the previous $kN-N$ dilution method, our error correction scheme completes the post-processing of oblivious key in this kind of QPQ protocols and makes them more practical for a real noisy channel. By our post-processing the error rate in the final oblivious key will be significantly decreased by the sacrifice on communication efficiency. For example, in the scenario where $N=10^5$, $k=7$, $p=0.25$, $e=3\%$, the error rate after our post-processing is 0.0008 instead of 0.1758 in the original $kN-N$ method without error correction. And at the same time, the user's privacy is still properly protected.

\vspace{1cm}
\textbf{Note.} When this work was finished we found that another error correction scheme for QPQ had been proposed by P. Chan et al in Ref.~\cite{otherEC}.

\section*{Acknowledgements}
This work is supported by NSFC (Grant Nos. 61300181, 61272057, 61202434, 61170270, 61100203, 61121061), Beijing Natural Science Foundation (Grant No. 4122054), Beijing Higher Education Young Elite Teacher Project (Grant Nos. YETP0475, YETP0477).



\begin{thebibliography}{ss3}
\bibitem{fn} In fact, different sub-key-extension operation (that is, different listing order when the users list the parities of any $k$ sub-key bits to obtain the extended sub-key) also just implies different orders of the row vectors in $G(s_i)$. So, our proof also works for the condition where different listing order is adapted.
\end{thebibliography}
\end{document}